\documentclass[12pt]{iopart}

\usepackage{txfonts,color}
\usepackage{graphicx}
\usepackage[sort,compress]{cite}
\usepackage{iopams}
\usepackage{setstack}
\usepackage{stmaryrd}
\usepackage{multirow,color}

\begin{document}

\title{Determinants of immediate price impacts at the trade level in an emerging order-driven market}

\author{Wei-Xing Zhou $^{1,2,3,4}$}

\address{$^1$ School of Business, East China University of Science and Technology, Shanghai 200237, People's Republic of China}
\address{$^2$ School of Science, East China University of Science and Technology, Shanghai 200237, People's Republic of China}
\address{$^3$ Research Center for Econophysics, East China University of Science and Technology, Shanghai 200237, People's Republic of China}

\ead{wxzhou@ecust.edu.cn (W.-X. Zhou)}

\begin{abstract}
The common wisdom argues that, in general, large trades cause large price changes, while small trades cause small price changes. However, for extremely large price changes, the trade size and news play a minor role, while the liquidity (especially price gaps on the limit order book) is a more influencing factor. Hence, there might be other influencing factors of immediate price impacts of trades. In this paper, through mechanical analysis of price variations before and after a trade of arbitrary size, we identify that the trade size, the bid-ask spread, the price gaps and the outstanding volumes at the bid and ask sides of the limit order book have impacts on the changes of prices. We propose two regression models to investigate the influences of these microscopic factors on the price impact of buyer-initiated partially filled trades, seller-initiated partially filled trades, buyer-initiated filled trades, and seller-initiated filled trades. We find that they have quantitatively similar explanation powers and these factors can account for up to 44\% of the price impacts. Large trade sizes, wide bid-ask spreads, high liquidity at the same side and low liquidity at the opposite side will cause a large price impact. We also find that the liquidity at the opposite side has a more influencing impact than the liquidity at the same side. Our results shed new lights on the determinants of immediate price impacts.
\end{abstract}

\submitto{\NJP}

\maketitle

{\color{blue}{\tableofcontents}}

\section{Introduction}
\label{s1:intro}

Econophysics is an interdiscipline of physics and economics, which applies methods and ideas in statistical physics to economic and financial systems \cite{Mantegna-Stanley-2000}. The distribution of assets is one of the most studied topics in econophysics \cite{Mantegna-Stanley-2000}, and in traditional finance as well \cite{Fama-1965-JB}, since it is the cornerstone of risk management and option pricing. The first work on this topic is usually traced back to Bachelier in 1900, where the Brownian motion is used to model the evolution of stock prices \cite{Bachelier-1900}. The Gaussian distribution of stock returns was rediscovered by Osborne in 1959 \cite{Osborne-1959a-OR}, which is a key assumption in the Black-Scholes option pricing model \cite{Black-Scholes-1973-JPE}. A pioneering work was done by Mandelbrot in 1963, who argued that the speculative price variations of cotton conform to the L\'evy distribution \cite{Mandelbrot-1963-JB}, which is a natural followup of his studies about income distributions \cite{Mandelbrot-1960-IER,Mandelbrot-1961-Em,Mandelbrot-1962-QJE,Mandelbrot-1963-JPE}. Using high-frequency data, Mantegna and Stanley found that a truncated L\'evy distribution is a better model for the S\&P 500 index returns \cite{Mantegna-Stanley-1995-Nature}. Further research showed that the returns have power-law tails with the exponent close to three, known as the inverse cubic law \cite{Gopikrishnan-Meyer-Amaral-Stanley-1998-EPJB,Liu-Gopikrishnan-Cizeau-Meyer-Peng-Stanley-1999-PRE,Gopikrishnan-Plerou-Amaral-Meyer-Stanley-1999-PRE,Plerou-Gopikrishnan-Amaral-Meyer-Stanley-1999-PRE}. A lot of empirical investigations have been conducted on financial returns at different time scales in different stock markets over different time periods, and the distributions are found to have power-law tails with different tail exponents at short time intervals
\cite{Makowiec-Gnacinski-2001-APP,Bertram-2004-PA,Yan-Zhang-Zhang-Tang-2005-PA,Qiu-Zheng-Ren-Trimper-2007-PA,Drozdz-Forczek-Kwapien-Oswicimka-Rak-2007-PA,Pan-Sinha-2007-EPL,Pan-Sinha-2008-PA,Tabak-Takami-Cajueiro-Petitiniga-2009-PA,Eryigit-Cukur-Eryigit-2009-PA,Jiang-Li-Cai-Wang-2009-PA,Queiros-2005-QF,Queiros-Moyano-deSouza-Tsallis-2007-EPJB,Zhang-Zhang-Kleinert-2007-PA,Gu-Chen-Zhou-2008a-PA,Fuentes-Gerig-Vicente-2009-PLoS1,Gerig-Vicente-Fuentes-2009-PRE}, which can be explained by the variational theory for turbulent signals
\cite{Castaing-Gagne-Hopfinger-1990-PD,Ghashghaie-Breymann-Peinke-Talkner-Dodge-1996-Nature}. In addition, it has been shown that the exponent values are universal for three mature markets that do not display variations with respect to market capitalization or industry sector \cite{Plerou-Stanley-2008-PRE,Stanley-Plerou-Gabaix-2008-PA}, which is however not the case for emerging markets \cite{Mu-Zhou-2010-PRE}.

So what causes the power-law tails in stock returns? A well-known adage in Wall Street states that it takes trading volume to move stock prices. The correlation between price fluctuation $r_{\Delta{t}}$ and trading volume $V_{\Delta{t}}$ has been extensively studied \cite{Karpoff-1987-JFQA}. Dominating evidence shows that the magnitude of price fluctuation is linearly correlated with the trading volume at different time scales $\Delta{t}$ from one minute to one month \cite{Wood-McInish-Ord-1985-JF,Jain-Joh-1988-JFQA,Ying-1966-Em,Epps-1977-JFQA,Harris-1987-JFQA,Gallant-Rossi-Tauchen-1992-RFS,Richardson-Sefcik-Thompson-1986-JFE,Rogalski-1978-RES,Saatcioglu-Starks-1998-IJF}:
\begin{equation}
 r_{\Delta{t}} = f(V_{\Delta{t}}) = \left\{
 \begin{array}{ccc}
   V_{\Delta{t}}/\lambda^+, & {\rm{for}} & V_{\Delta{t}}>0\\
   V_{\Delta{t}}/\lambda^-, & {\rm{for}} & V_{\Delta{t}}<0\\
 \end{array}
 \right.
 \label{Eq:r:kV}
\end{equation}
where $f$ is the price impact function and $\lambda$ is a measure of market liquidity. In addition, the price-volume relation is asymmetric such that $\lambda^+>\lambda^->0$, that is, the price impact of a selling volume is larger than a buying volume of the same size \cite{Karpoff-1987-JFQA}. At the transaction level, theoretical analyses and empirical evidence shows that the price impact function is nonlinear \cite{Loeb-1983-FAJ,Perold-Salomon-1991-FAJ,Zhang-1999-PA,Farmer-2002-ICC,Almgren-2003-AMF,Gabaix-Gopikrishnan-Plerou-Stanley-2003-Nature,Gabaix-Gopikrishnan-Plerou-Stanley-2003-PA,Lillo-Farmer-Mantegna-2003-Nature,Lim-Coggins-2005-QF,Farmer-Lillo-2004-QF,Plerou-Gopikrishnan-Gabaix-Stanley-2004-QF,Gabaix-Gopikrishnan-Plerou-Stanley-2006-QJE,Gabaix-Gopikrishnan-Plerou-Stanley-2007-JEEA,Zhou-2012-QF}.

Concerning the volume-return relation at the transaction level, we are interested in the determination of the immediate price impact $r_{k,t+1}$ of trade size $\omega_{k,t}$ for a given stock $k$. The immediate price impact can be calculated as follows:
\begin{equation}
 r_{k,t+1} = \frac{p_{k,t+1}-p_{k,t}}{p_{k,t}} = \frac{a_{k,t+1}+b_{k,t+1}-a_{k,t}-b_{k,t}}{a_{k,t}+b_{k,t}},
 \label{Eq:Ret}
\end{equation}
where $p_{k,t}$, $a_{k,t}$, and $b_{k,t}$ are the mid-price, the best ask price and the best bid price right before the trade, and $p_{k,t+1}$, $a_{k,t+1}$, and $b_{k,t+1}$ are the mid-price, the best ask price and the best bid price immediately after the trade. Using the Trade and Quote (TAQ) database, Lillo et al unveiled a master curve for price impact function such that the data collapse onto a single curve \cite{Lillo-Farmer-Mantegna-2003-Nature}:
\begin{equation}\label{Eq:LFM}
    r\left(\frac{\omega}{\langle\omega\rangle},C\right)
    =C^{-\gamma}g\left(\frac{\omega}{\langle\omega\rangle}\frac{1}{C^\delta}\right)~,
\end{equation}
where $r$ is the shift of logarithmic mid-quote prices right before and after a trade of size $\omega$ occurs, $\langle\omega\rangle$ is the average volume per trade for each stock, $C$ is the stock capitalization, $\gamma$ and $\delta$ are two scaling parameters, and $g(x)$ is found to be a concave function and has a power-law form for large $\omega$. Lim and Coggins performed a similar analysis on the Australian Stock Exchange and a similar scaling was obtained, where the return and trade size were defined in a slightly different way by taking into account the intraday pattern effect \cite{Lim-Coggins-2005-QF}. Using order flow data in the Chinese market, Zhou found that trades resulting from filled and partially filled limit orders have very different price impacts \cite{Zhou-2012-QF}. The price impact of trades from partially filled orders is constant when the volume is not too large, while that of filled orders shows a power-law behavior. When the returns and the trade sizes are normalized by their stock-dependent averages, capitalization-independent scaling laws emerge for both buyer-initiated and seller initiated trades from filled orders \cite{Zhou-2012-QF}:
\begin{equation}\label{Eq:Zhou}
    \frac{r_k}{\langle{r_k}\rangle} = A_k\left(\frac{\omega_k}{\langle\omega_k\rangle}\right)^\alpha,
\end{equation}
where $\alpha \approx 2/3$ and the prefactor $A_k=A$ is independent of stock $k$ and its capitalization.

In the aforementioned studies, the trading volume or trade size is considered as the solo driving factor of price fluctuations. For larger price fluctuations over a fixed clock time scale $\Delta{t}$, more determinants of price impact have been investigated. Farmer et al investigated the transaction data of 16 companies traded on the London Stock Exchange (LSE) in the 4-year period 1999-2002 and found that large price fluctuations are essentially independent of the volume of order, but driven by liquidity fluctuations characterized by the gaps between the first few price levels on the opposite limit order book \cite{Farmer-Gillemot-Lillo-Mike-Sen-2004-QF}. Using the TAQ data for the year 1997 and the order book data from the Island ECN for the year 2002, Weber and Rosenow argued that a large trading volume alone is not sufficient to explain large price changes and a lack of liquidity is a necessary prerequisite for the occurrence of large price fluctuations \cite{Weber-Rosenow-2006-QF}. Joulin et al analyzed the one-minute data of 163 USA stocks and found that news and trading volume play a minor role in causing large price changes, which is thus conjectured to be caused by the vanishing of liquidity \cite{Joulin-Lefevre-Grunberg-Bouchaud-2008-Wilmott}. N{\ae}s and Skjeltorp studied the order flow data from the Oslo Stock Exchange in Norway and found that price fluctuations are positively correlated with trade number, a component of trading volume, and negatively correlated with different liquidity measures \cite{Naes-Skjeltorp-2006-JFinM}.

In this work, we perform research on the determinants of immediate price impact defined in equation (\ref{Eq:Ret}) in an emerging stock market. This research is designed to have many differences from the literature. We focus on the immediate price impact after a transaction occurs and investigate filled and partially filled orders separately. We use three measures for liquidity based on the bid-ask spread, the standing volume on the limit order book, and the gaps between successive price levels.

The motivation of this research is the following. As we briefly reviewed above, the determinants of immediate price impacts have not been investigated for the Chinese stock market, which is a representative emerging market. Our analysis here goes further by investigate separately filled and partially filled orders since these two kinds of orders have nontrivially very different behaviors \cite{Zhou-2012-QF}. We expect that the influencing factors will have different impacts on the price movement. This research is empirical and not intended for practical use. The new empirical results obtained in this work thus deepens our understanding of the dynamics of this specific market. For a given order, when all the independent variables are known, we can calculate the immediate price impact for sure. However, this treatment for individual orders can not be generalized to the macroscopic level for a population of orders. It is thus necessary to perform regression analysis on the data.

The rest of this paper is organized as follows. Section \ref{S1:Data} describes the data used in our study. Section \ref{S1:Model} discusses the possible determinants of immediate price impact and specifies the model. The results are shown in section \ref{S1:IndStocks} for individual stocks and in section \ref{S1:AllStocks} for all stocks. We summarize our findings in section \ref{S1:Conc}.

\section{Data sets}
\label{S1:Data}

We use the order flow data of 23 A-share stocks traded on the Shenzhen Stock Exchange (SZSE) of China in 2003. The A-shares are common stocks issued by mainland Chinese companies, subscribed and traded in Chinese currency {\em{Renminbi}}, listed on mainland Chinese stock exchanges, bought and sold by Chinese nationals and approved foreign investors. The A-share market was open only to domestic investors in 2003. Note that the Chinese stock market is the largest emerging market in the world and became the second largest stock market after the USA market in 2009. Our sample stocks were part of the 40 constituent stocks included in the Shenshen Stock Exchange component index in 2003.

\begin{table}[t!]
 \centering
 \caption{\label{Tb:BasicStat}Basic statistics for the 23 SZSE stocks. The columns give the codes of stocks, the average returns  ($\langle{r_{\rm{PB}}}\rangle$, $\langle{r_{\rm{PS}}}\rangle$, $\langle{r_{\rm{FB}}}\rangle$, and $\langle{r_{\rm{FS}}}\rangle$, multiplied by 1000), the numbers of trades ($N$, thousand shares), the fractions of partially filled orders ($F$), and the industrial sectors.}
\bigskip
\begin{tabular}{cccccr@{.}lcl}
  \hline\hline
   Code
        & $\langle{r_{\rm{PB}}}\rangle$ & $\langle{r_{\rm{PS}}}\rangle$
        & $\langle{r_{\rm{FB}}}\rangle$ & $\langle{r_{\rm{FS}}}\rangle$ &\multicolumn{2}{c}{$N$} & $F$  & Industry\\  \hline
000001 &  1.19 & -1.21 &  0.03 & -0.05 &   889&7  & 0.07  &           Financials\\
000002 &  1.55 & -1.54 &  0.04 & -0.05 &   509&4  & 0.06  &          Real estate\\
000009 &  2.23 & -2.25 &  0.05 & -0.06 &   448&0  & 0.08  &        Conglomerates\\
000012 &  1.63 & -1.61 &  0.08 & -0.10 &   290&4  & 0.14  &  Metals \& Nonmetals\\
000016 &  1.86 & -1.87 &  0.08 & -0.10 &   188&6  & 0.15  &          Electronics\\
000021 &  1.35 & -1.37 &  0.07 & -0.09 &   411&6  & 0.12  &          Electronics\\
000024 &  1.65 & -1.63 &  0.09 & -0.09 &   133&6  & 0.16  &          Real estate\\
000027 &  1.63 & -1.63 &  0.06 & -0.05 &   313&9  & 0.08  &            Utilities\\
000063 &  1.27 & -1.24 &  0.07 & -0.06 &   265&5  & 0.10  &                   IT\\
000066 &  1.57 & -1.59 &  0.08 & -0.09 &   277&7  & 0.14  &          Electronics\\
000088 &  1.63 & -1.60 &  0.10 & -0.09 &    97&2  & 0.13  &       Transportation\\
000089 &  1.63 & -1.64 &  0.06 & -0.07 &   189&1  & 0.11  &       Transportation\\
000406 &  1.56 & -1.58 &  0.05 & -0.07 &   271&4  & 0.13  &       Petrochemicals\\
000429 &  2.35 & -2.37 &  0.08 & -0.09 &   117&4  & 0.14  &       Transportation\\
000488 &  1.72 & -1.62 &  0.14 & -0.17 &   120&1  & 0.21  &    Paper \& Printing\\
000539 &  1.88 & -1.85 &  0.10 & -0.14 &   114&7  & 0.14  &            Utilities\\
000541 &  1.56 & -1.54 &  0.09 & -0.09 &    68&7  & 0.18  &          Electronics\\
000550 &  1.60 & -1.59 &  0.08 & -0.09 &   346&7  & 0.12  &        Manufacturing\\
000581 &  1.84 & -1.80 &  0.10 & -0.10 &    94&0  & 0.17  &        Manufacturing\\
000625 &  1.60 & -1.61 &  0.08 & -0.10 &   397&6  & 0.12  &        Manufacturing\\
000709 &  2.02 & -2.04 &  0.04 & -0.05 &   207&8  & 0.10  &  Metals \& Nonmetals\\
000720 &  0.98 & -1.15 &  0.05 & -0.08 &   132&2  & 0.16  &            Utilities\\
000778 &  1.38 & -1.33 &  0.07 & -0.07 &   157&3  & 0.15  &        Manufacturing\\
  \hline\hline
\end{tabular}
\end{table}

The SZSE is open for trading from Monday to Friday except the public holidays and other dates as announced by the China Securities Regulatory Commission. On each trading day, the market opens at 9:15 am and entered the opening call auction till 9:25 am, during which the trading system accepts order submission and cancelation, and all matched transactions are executed at 9:25 am. It is followed by a cooling period from 9:25 am to 9:30. During cooling periods, the Exchange is open to orders routing from members, but does not process orders or process cancelation of orders. The information released to trading terminals also does not change during cooling periods. All matched transactions are executed at the end of cooling periods. The continuous double auction operates from 9:30 to 11:30 and 13:00 to 15:00 and transaction occurs based on automatic one to one matching due to price-time priority. The market freezes in the time period from 11:30 am to 13:00 pm and no actions could be taken. At the daily closure of the market, unexecuted orders will be automatically removed. Only the trades during the continuous double auction are considered in this work.

Following reference \cite{Biais-Hillion-Spatt-1995-JF}, the trades are differentiated into four types according to their directions (whether a trade is seller-initiated or buyer-initiated) and aggressiveness: buyer-initiated partially filled (PB) trades resulting from partially filled buy orders, seller-initiated partially filled (PS) trades resulting from partially filled sell orders, buyer-initiated filled (FB) trades resulting from filled buy orders, and seller-initiated filled (FS) trades resulting from filled sell orders. The basic statistics for the 23 SZSE stocks are presented in table \ref{Tb:BasicStat} \cite{Zhou-2012-QF}. We note that the 23 stocks investigated in this work are representative, which belong to a variety of industry sectors as shown in table \ref{Tb:BasicStat}.

Several interesting features can be observed from table \ref{Tb:BasicStat}. First, the majority of trades are resulted from filled orders, which means that these investors are subjectively more impatient. Second, we find that the immediate price impacts of buyer-initiated and seller-initiated trades are roughly symmetric since
\begin{equation}
 \langle{r_{\rm{PB}}}\rangle \approx -\langle{r_{\rm{PS}}}\rangle~~~
 {\rm{and}}~~~
 \langle{r_{\rm{FB}}}\rangle \approx -\langle{r_{\rm{FS}}}\rangle
 \label{Eq:rr:BS:symetric}
\end{equation}
for each stock. Third, partially filled trades have much larger impact on the price than filled trades \cite{Zhou-2012-QF}:
\begin{equation}
 \langle{r_{\rm{PB}}}\rangle \gg \langle{r_{\rm{FB}}}\rangle~~~
 {\rm{and}}~~~-\langle{r_{\rm{PS}}}\rangle\gg -\langle{r_{\rm{FS}}}\rangle~,
 \label{Eq:rr}
\end{equation}
which means that, on average, partially filled trades are much more aggressive than filled trades, which is confirmed by the fact that partially filled trades have larger sizes. A mechanical illustration of the third feature will be shown in section \ref{S1:Model}.

\section{Model specification}
\label{S1:Model}

\begin{figure}[tb]
 \centering
 \includegraphics[width=7.5cm]{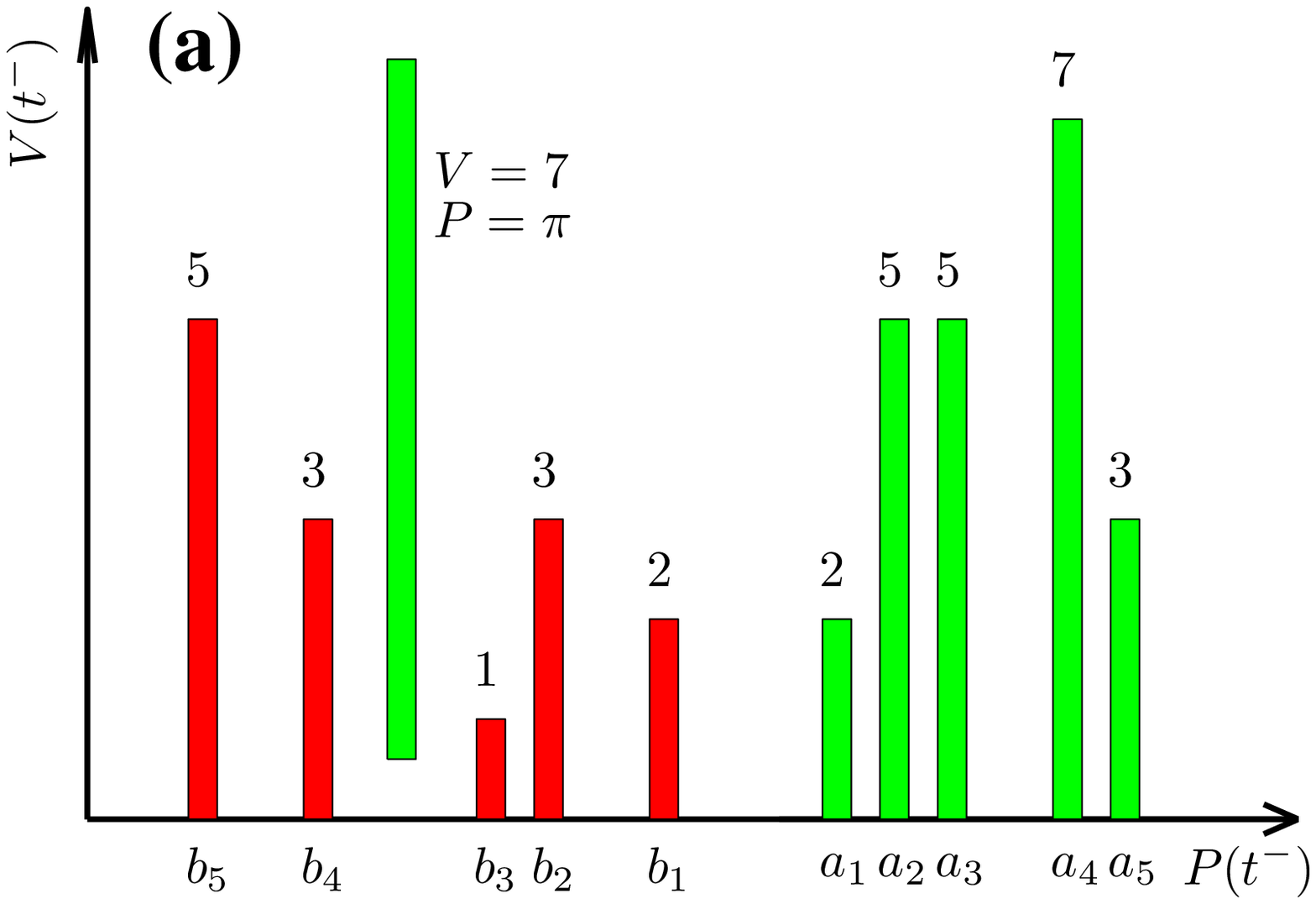}
 \includegraphics[width=7.5cm]{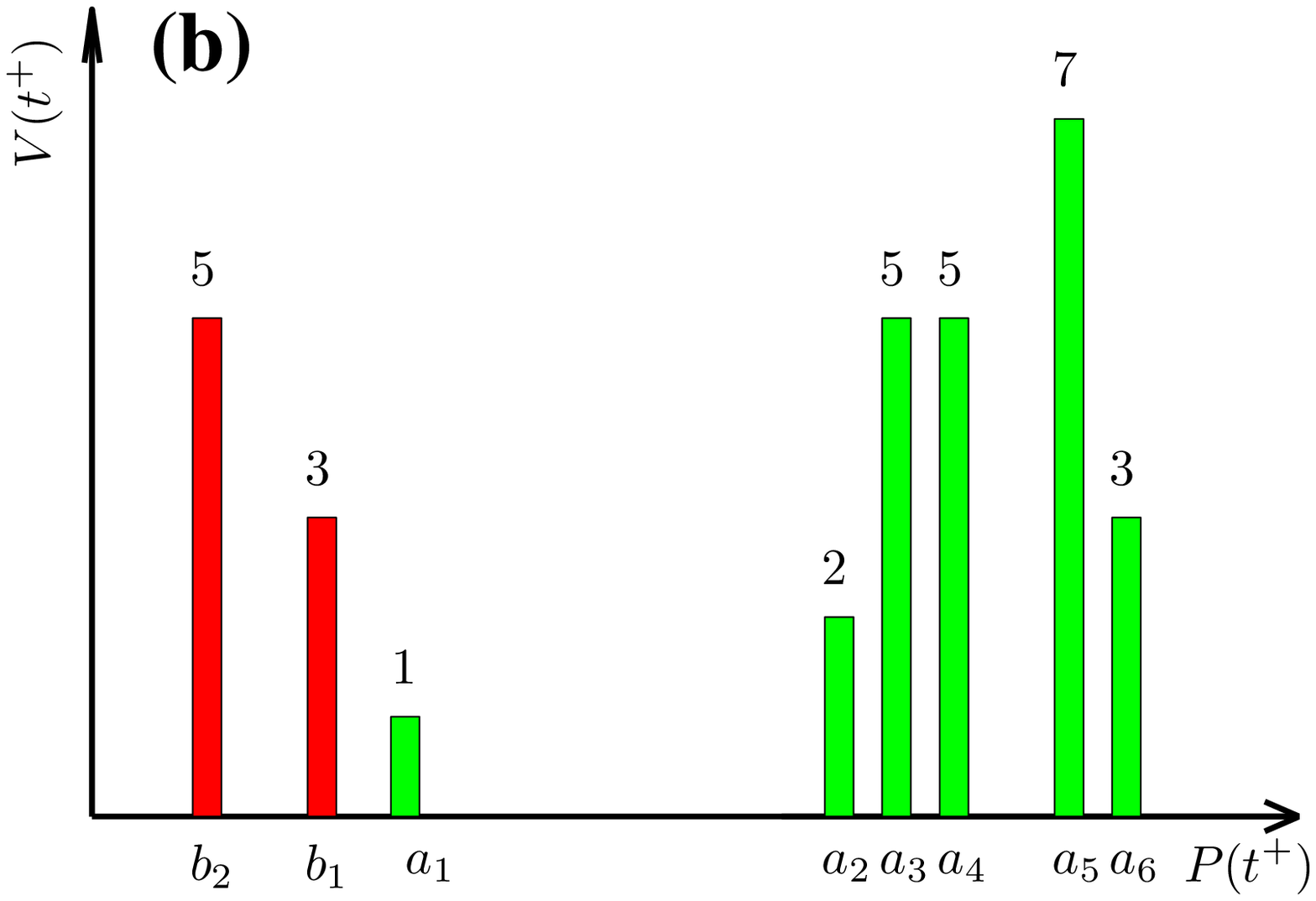}
 \includegraphics[width=7.5cm]{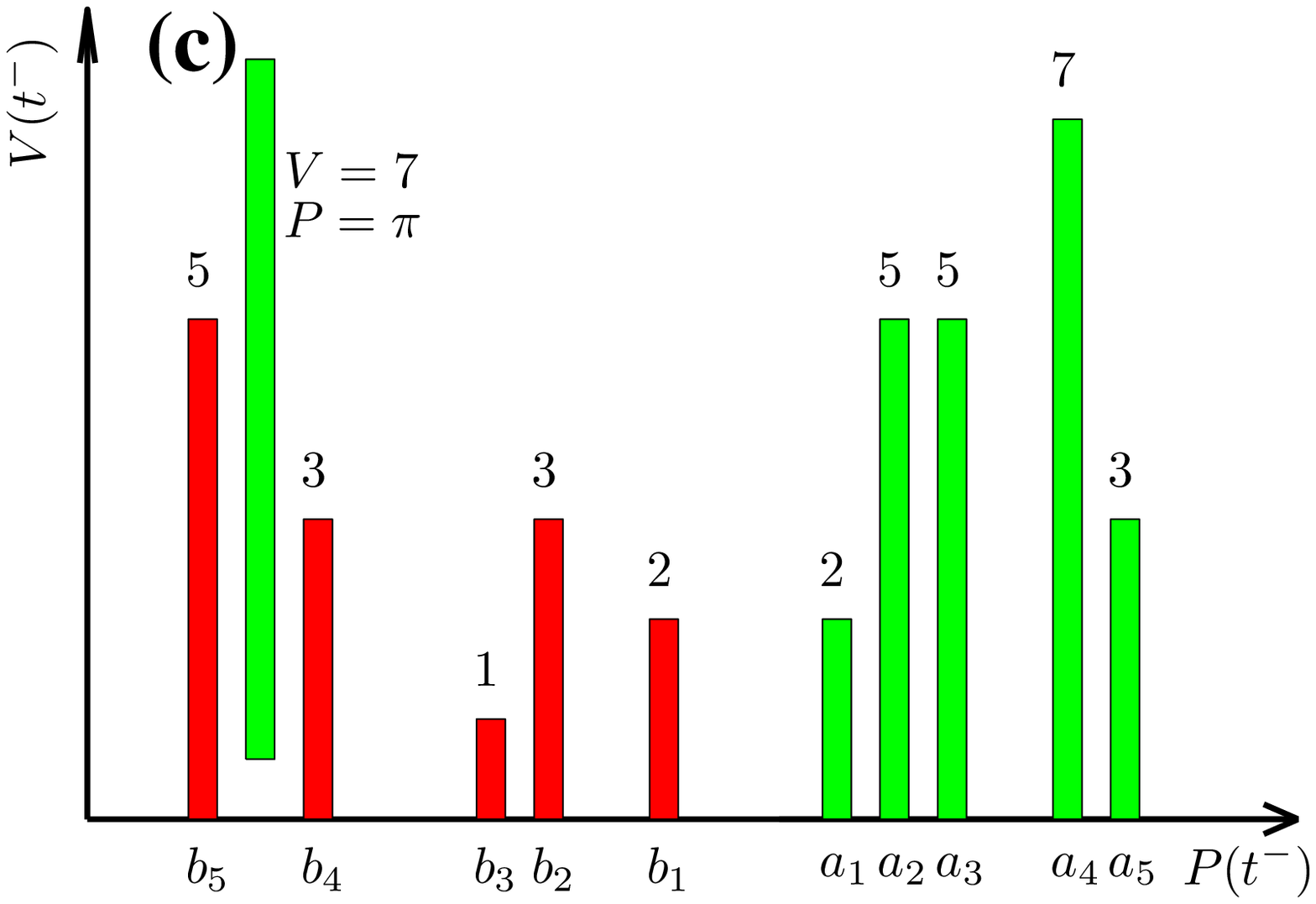}
 \includegraphics[width=7.5cm]{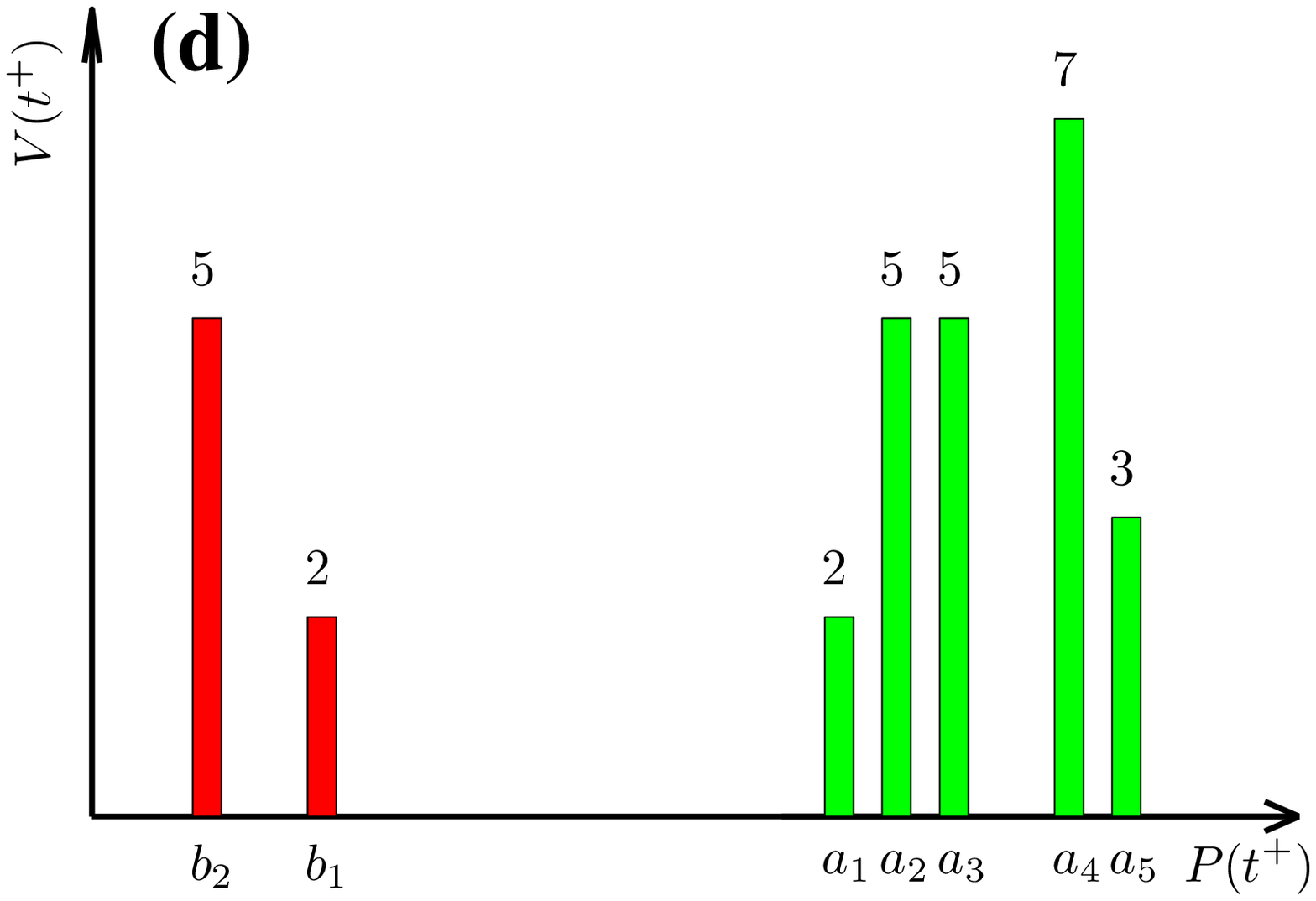}
 \caption{\label{Fig:LOB:t} Schematic illustration of price formation caused by seller-initiated partially filled orders (a-b) and filled orders (c-d). The numbers above the bars are the volumes at the corresponding levels.}
\end{figure}

In order to determine the influence factors of immediate price impact, we show in figure \ref{Fig:LOB:t} the evolution of the limit order book and the price formation process of seller-initiated partially filled orders and filled orders.

Consider the situation that a sell order with price $P=\pi$ and size $V=7$ is submitted at time $t$. Figure \ref{Fig:LOB:t}(a) shows the first five price levels of both sides of the limit order book right before the placement of an sell order, in which the ask prices, the bid prices and their volumes are denoted by $a_i(t^-)$, $b_i(t^-)$, $V^A_i(t^-)$, and $V^B_i(t^-)$, respectively. Assume that $b_4(t^-)<\pi<b_3(t^-)$. At time $t$, the buy limit orders on the first three price levels are executed, whose trading volume is $\omega_t=V^B_1(t^-)+V^B_2(t^-)+V^B_3(t^-)=6$. The unexecuted part of the sell order, whose quantity is $V-\omega_t=1$, is stored on the sell limit order book at time $t^+$, as shown in figure \ref{Fig:LOB:t}(b), where we have $a_1(t^+)=\pi$, $a_{i+1}(t^+)=a_i(t^-)$ for $i=1,2,\cdots$, and $b_i(t^+)=b_{i+3}(t^-)$. More generally, if we have
\begin{equation}
  b_{n+1}(t^-)<\pi<b_{n}(t^-)~~~{\rm{and}}~~~\sum_{i=1}^{n} V^B_i(t^-)<V,
  \label{Eq:LOB:PS:Cond}
\end{equation}
then the sell order eats buy limit orders waiting on the order book and the new best ask and bid prices are $b_1(t^+)=b_{n+1}(t^-)$ and $a_1(t^+)=\pi< a_1(t^-)$. According to equation (\ref{Eq:Ret}), the immediate price impact is
\begin{eqnarray}
 r_{\rm{PS}}&=&\frac{\pi+b_{n+1}(t^-)-a_1(t^-)-b_1(t^-)}{a_1(t^-)+b_1(t^-)}\\
            &=&-\frac{a_1(t^-)-b_1(t^-)}{a_1(t^-)+b_1(t^-)}-\frac{b_1(t^-)-b_{n+1}(t^-)}{a_1(t^-)+b_1(t^-)}-\frac{b_1(t^-)-\pi}{a_1(t^-)+b_1(t^-)}.
 \label{Eq:r:PS}
\end{eqnarray}
The first term in the right-hand side of equation (\ref{Eq:r:PS}) indicates that the immediate price impact $r_{\rm{PS}}$ of partially filled orders is negatively proportional to the bid-ask spread $S(t^-)$. The second term implies that $r_{\rm{PS}}$ is negatively proportional to the first $n$ price gaps
\begin{equation}
 G_i^B(t^-)=b_i(t^-)-b_{i+1}(t^-),~~~ i=1,\cdots,n,
 \label{Eq:GiB}
\end{equation}
and correlated with the volumes $V_i^B(t^-)$ on the first $n$ levels $i=1,2,\cdots,n$. The correlation between $r_{\rm{PS}}$ and $V_i^B(t^-)$ is due to the fact that the value of $n$ is determined by the order book shape according to equation (\ref{Eq:LOB:PS:Cond}). The third term also indicates that $r_{\rm{PS}}$ is correlated with $G_i^B(t^-)$ on the first $n$ levels $i=1,2,\cdots,n$. Note that the correlation between $r_{PS}$ and $\omega$ can also be inferred from equation (\ref{Eq:LOB:PS:Cond}). Taking into consideration the finding expressed in equation (\ref{Eq:Zhou}) \cite{Zhou-2012-QF}, we can propose the following model:
\begin{equation}
 r_{\rm{PS}}(t) = a_0 + a \left[\omega(t^-)\right]^\alpha  + b S(t^-)
           + \sum_{i=1}^{n} d_i \left[{V^B_i(t^-)}\right]^\beta
           + \sum_{i=1}^{n} f_i G^B_i(t^-)
           + u_t,
 \label{Eq:Pi:PS}
\end{equation}
where the coefficients $a$, $b$ and $f_i$ are expected to be negative and the $d_i$ coefficients are expected to be positive.  Similarly, for partially filled buy orders, we can propose the following model:
\begin{equation}
 r_{\rm{PB}}(t) = a_0 + a \left[\omega(t^-)\right]^\alpha  + b S(t^-)
           + \sum_{i=1}^{n} c_i \left[{V^A_i(t^-)}\right]^\beta
           + \sum_{i=1}^{n} e_i G^A_i(t^-)
           + u_t,
 \label{Eq:Pi:PB}
\end{equation}
where the coefficients $a$, $b$ and $e_i$ are expected to be positive and the $c_i$ coefficients are expected to be negative.

Panels (c) and (d) of figure \ref{Fig:LOB:t} illustrate the evolution of the limit order book around a filled sell order. If the sell order is filled such that
\begin{equation}
  \pi\leqslant b_{n}(t^-)~~~{\rm{and}}~~~\sum_{i=1}^{n} V^B_i(t^-) = V,
  \label{Eq:LOB:FS:Cond1}
\end{equation}
or
\begin{equation}
  \pi\leqslant b_{n+1}(t^-)~~~{\rm{and}}~~~\sum_{i=1}^{n} V^B_i(t^-)< V < \sum_{i=1}^{n+1} V^B_i(t^-),
  \label{Eq:LOB:FS:Cond2}
\end{equation}
then the new best ask and bid prices are $b_1(t^+)=b_{n+1}(t^-)$ and $a_1(t^+)=a_1(t^-)$. Hence, the immediate price impact is
\begin{equation}
 r_{\rm{FS}} = \frac{b_{n+1}(t^-)-b_1(t^-)}{a_1(t^-)+b_1(t^-)}.
 \label{Eq:r:FS}
\end{equation}
The equation implies that $r_{\rm{FS}}$ is negatively proportional to the first $n$ price gaps $G_i^B(t^-)$ and correlated with the volumes $V_i^B(t^-)$ on the first $n$ levels $i=1,2,\cdots,n$, but independent of the bid-ask spread.
We can thus propose the following model:
\begin{equation}
 r_{\rm{FS}}(t) = a_0 + a \left[\omega(t^-)\right]^\alpha
           + \sum_{i=1}^{n} d_i \left[{V^B_i(t^-)}\right]^\beta
           + \sum_{i=1}^{n} f_i G^B_i(t^-)
           + u_t,
 \label{Eq:Pi:FS}
\end{equation}
where the coefficients $a$ and $f_i$ are expected to be negative and the $d_i$ coefficients are expected to be positive.  Similarly, for filled buy orders, we can propose the following model:
\begin{equation}
 r_{\rm{FB}}(t) = a_0 + a \left[\omega(t^-)\right]^\alpha
           + \sum_{i=1}^{n} c_i \left[{V^A_i(t^-)}\right]^\beta
           + \sum_{i=1}^{n} e_i G^A_i(t^-)
           + u_t,
 \label{Eq:Pi:FB}
\end{equation}
where the coefficients $a$ and $e_i$ are expected to be positive and the $c_i$ coefficients are expected to be negative.

It follows immediately from equations (\ref{Eq:r:PS}) and (\ref{Eq:r:FS}) that
\begin{equation}
 r_{\rm{PS}}-r_{\rm{FS}} = -\frac{a_1(t^-)-\pi}{a_1(t^-)+b_1(t^-)}
                         = -\frac{a_1(t^-)-b_1(t^-)}{a_1(t^-)+b_1(t^-)} -\frac{b_1(t^-)-\pi}{a_1(t^-)+b_1(t^-)}.
 \label{Eq:r:PS:FS}
\end{equation}
whose absolute value is no less than half of the relative bid-ask spread $[a_1(t^-)-b_1(t^-)]/[a_1(t^-)+b_1(t^-)]$. Our data show that $91.05\%$ of the FS trades have $r_{\rm{FS}}=0$, and all price movements caused by the PS trades are negative, which is not surprising in view of the limit order book mechanism. These considerations explain why PS trades have much larger price impacts than FS trades. This analysis applies for buyer-initiated trades as well. We note that $89.42\%$ of the FB trades have $r_{\rm{FB}}=0$.

There are possible missing determinants of immediate price impact in the above discussions. One factor is the liquidity in the same side of the limit order book for executed buy or sell orders, since the order book shape and the price gaps on the buy and sell sides are statistically symmetric \cite{Gu-Chen-Zhou-2008c-PA}. Another factor is the intraday patterns found in many financial quantities including the volatility, the bid-ask spread, and the trading volume. Hence, the immediate price impact of each type of trades for stock $k$ can be modeled as follows:
\begin{equation}
r_{k,t+1} =  f(\omega_{k,t}, S_{k,t}, V^A_{k,i=1:L,t}, V^B_{k,i=1:L,t}, G^A_{k,i=1:L,t}, G^B_{k,i=1:L,t}, D_{i=1:23,t}) + u_t,
 \label{Eq:Pi:IP:Model:1}
\end{equation}
where $D_{i=1:23,t}$ are dummy variables characterizing the intraday pattern in a resolution of ten minutes. We consider two types of models. The first type is the following:
\begin{eqnarray}
\frac{r_{k,t+1}}{\langle{r_k}\rangle}
          &=&
              a_0 + a \left(\frac{\omega_{k,t}}{\langle{\omega_k}\rangle}\right)^\alpha  + b S_{k,t}   %
           + \sum_{i=1}^{L} c_i \left(\frac{V^A_{k,i,t}}{\langle{\omega_k}\rangle}\right)^\beta %
           + \sum_{i=1}^{L} d_i \left(\frac{V^B_{k,i,t}}{\langle{\omega_k}\rangle}\right)^\beta %
           \nonumber\\&&
           + \sum_{i=1}^{L} e_i \frac{G^A_{k,i,t}}{|\langle{r_k}\rangle|}
           + \sum_{i=1}^{L} f_i \frac{G^B_{k,i,t}}{|\langle{r_k}\rangle|}
           + \sum_{i=1}^{23} g_i D_{i,t} + u_t,
    \label{Eq:Pi:IP:Model:PL}
\end{eqnarray}
where the variables are normalized so that we can put together all the data of different stocks in the cross-sectional analysis in section \ref{S1:AllStocks}, following reference \cite{Zhou-2012-QF}. The second model is expressed as follows:
\begin{eqnarray}
\frac{r_{k,t+1}}{\langle{r_k}\rangle}
          &=&
              a_0 + a \left(\frac{\omega_{k,t}}{\langle{\omega_k}\rangle}\right)^\alpha  + b S_{k,t}   %
           + \sum_{i=1}^{L} c_i \ln\left(\frac{V^A_{k,i,t}}{\langle{\omega_k}\rangle}\right) %
           + \sum_{i=1}^{L} d_i \ln\left(\frac{V^B_{k,i,t}}{\langle{\omega_k}\rangle}\right) %
           \nonumber\\&&
           + \sum_{i=1}^{L} e_i \frac{G^A_{k,i,t}}{|\langle{r_k}\rangle|}
           + \sum_{i=1}^{L} f_i \frac{G^B_{k,i,t}}{|\langle{r_k}\rangle|}
           + \sum_{i=1}^{23} g_i D_{i,t} + u_t.
    \label{Eq:Pi:IP:Model:LN}
\end{eqnarray}
For simplicity, we term the model expressed in equation (\ref{Eq:Pi:IP:Model:PL}) as the {\it{power-law model}} and the model in equation (\ref{Eq:Pi:IP:Model:LN}) as the {\it{logarithmic model}}.

Since the size of the data is huge, we preprocess the data for each stock before calibrating the two models expressed in equations (\ref{Eq:Pi:IP:Model:PL}) and (\ref{Eq:Pi:IP:Model:LN}) to reduce the computational time. For each stock, the average values of $r_{k,t+1}$, $S_{k,t}$, $V^A_{k,i=1:L,t}$, $V^B_{k,i=1:L,t}$, $G^A_{k,i=1:L,t}$, and $G^B_{k,i=1:L,t}$ of the trades with the same size $\omega_{k,t}$ are used. The parameters $a_0$, $a$, $b$, $c_i$, $d_i$, $e_i$, and $f_i$ are determined by minimizing the following objective function:
\begin{equation}
 Q(\alpha,\beta; a_0, a, b, c_i, d_i, e_i, f_i) = \sum_t u_t^2.
 \label{Eq:ObjFun}
\end{equation}
This least-squares estimation is equivalent to the maximum likelihood estimation only if the measurement errors are independent and identically distributed Gaussian \cite{Press-Teukolsky-Vetterling-Flannery-1996}. However, this condition may not hold. For stock price movements, it is known that both limit orders for liquidity provision and market orders for liquidity taking have long memory properties \cite{Bouchaud-Gefen-Potters-Wyart-2004-QF,Lillo-Farmer-2004-SNDE,Gu-Zhou-2009-EPL}, and these two effects usually reconcile to produce diffusive price fluctuations \cite{Bouchaud-Gefen-Potters-Wyart-2004-QF,Lillo-Farmer-2004-SNDE,Bouchaud-2005-Chaos}, be the price impact temporal or permanent \cite{Farmer-Gerig-Lillo-Mike-2006-QF,Bouchaud-Kockelkoren-Potters-2006-QF}. On the other hand, we do not know a priori the distribution of the measurement errors. Therefore, we use this least-squares estimation in this work. We note that it might underestimate the $p$-values, which however should not change the quantitative conclusions.

In order to achieve the global optimization of the power-law model, we scan $\alpha$ and $\beta$ in a two-dimensional grid from 0.05 to 0.95 with a spacing step of 0.05. For each pair of $\alpha$ and $\beta$, we can perform a linear least-squares regression to determine the local optimal parameters and the adjusted R-square, $R^2$-adj. For the logarithmic model, we need to scan $\alpha$ only. The set of parameters that maximizing the $R^2$-adj is considered as the globally optimal solution.

\section{Results for individual stocks}
\label{S1:IndStocks}

\subsection{The case of five levels of liquidity $L=5$}
\label{S2:IndStock:L5}

For each stock, we have calibrated the power-law model in equation (\ref{Eq:Pi:IP:Model:PL}) with $L=5$ for the four types of trades, {\it{i.e.}}, buyer-initiated partially filled (PB) trades, seller-initiated partially filled (PS) trades, buyer-initiated filled (FB) trades, and seller-initiated filled (FS) trades. For each type of trades, the optimal values of $\alpha$ and $\beta$ are obtained by maximizing the adjusted R-square ($R^2$-adj).

Taking stock 000001 as an example, we have $\alpha=0.25$ and $\beta=0.15$ for PB trades, $\alpha=0.30$ and $\beta=0.20$ for PS trades, $\alpha=0.55$ and $\beta=0.10$ for FB trades, and $\alpha=0.45$ and $\beta=0.05$ for FS trades. The regression results are shown in Table \ref{Tb:Model:PL:L5:000001}. The F-tests show that the model is significant for all the four types of trades and the explanatory power of the variable is very high. For all four types of trades, the normalized immediate price impact $r/\langle{r}\rangle$ is positively correlated with the normalized trade size $\omega$ and the normalized bid-ask spread $S$. However, the relation between $r$ and $S$ is insignificant for the FS trades, which is consistent with the simple explanation in equation (\ref{Eq:r:FS}). For each type of trades, roughly speaking, the normalized immediate price impact is positively correlated with the standing volumes on the same side of the limit order book and the price gaps on the opposite side, and negatively correlated with the standing volumes on the opposite side and the price gaps on the same side. In other words, the magnitude of the price impact is large if the liquidity is high on the same side (large standing volumes and narrow price gaps) or if the liquidity is low on the opposite side (small standing volumes and wide price gaps). A closer scrutiny unveils that the liquidity on the same side of the limit order book has a less influence on the price impact. For instance, all coefficients for $V_i^A$ and $G_i^A$ are statistically insignificant for FS trades.

\begin{table}[tb]
 \caption{\label{Tb:Model:PL:L5:000001} Calibration of the power-law model in (\ref{Eq:Pi:IP:Model:PL}) with $L=5$ for stock 000001. We obtain that $\alpha=0.25$ and $\beta=0.15$ for buyer-initiated partially filled (PB) trades, $\alpha=0.30$ and $\beta=0.20$ for seller-initiated partially filled (PS) trades, $\alpha=0.55$ and $\beta=0.10$ for buyer-initiated filled (FB) trades, and $\alpha=0.45$ and $\beta=0.05$ for seller-initiated filled (FS) trades. The numbers in the parentheses are the $p$-values. The estimates of the coefficients in red indicate that the associated variables are significant at the 5\% level.}
 \medskip
 \centering
  \begin{tabular}{cr@{ }lr@{ }lr@{ }lr@{ }l cc}
  \hline \hline
    Variable & \multicolumn{2}{c}{PB} & \multicolumn{2}{c}{PS} & \multicolumn{2}{c}{FB} & \multicolumn{2}{c}{FS} \\\hline %
    $R^2$-adj & \multicolumn{2}{c}{0.475} & \multicolumn{2}{c}{0.501} & \multicolumn{2}{c}{0.444} & \multicolumn{2}{c}{0.468} \\ %
    $p$-value & \multicolumn{2}{c}{0.000} & \multicolumn{2}{c}{0.000} & \multicolumn{2}{c}{0.000} & \multicolumn{2}{c}{0.000} \\ %
  \hline
  $a_0$ &  {\color{red}{0.98}}&(0.000) &  {\color{red}{0.51}}&(0.000) & {\color{red}{31.59}}&(0.000) & {\color{red}{30.76}}&(0.000) \\%
  $a$ for $\omega$ &  {\color{red}{1.15}}&(0.000) &  {\color{red}{1.06}}&(0.000) &  {\color{red}{1.85}}&(0.000) &  {\color{red}{2.86}}&(0.000) \\%
  $b$ for $S$ & {\color{red}{20.01}}&(0.000) & {\color{red}{31.26}}&(0.000) & {\color{red}{227.19}}&(0.000) &  3.38&(0.635) \\%
  \hline
  $c_1$ for $V^A_1$ & {\color{red}{-1.82}}&(0.000) &  {\color{red}{0.08}}&(0.000) & {\color{red}{-30.83}}&(0.000) &  1.08&(0.072) \\%
  $c_2$ for $V^A_2$ & {\color{red}{-0.28}}&(0.000) &  0.01&(0.405) & {\color{red}{-8.73}}&(0.000) &  1.26&(0.072) \\%
  $c_3$ for $V^A_3$ &  0.01&(0.585) & -0.03&(0.065) & -1.80&(0.042) &  1.35&(0.063) \\%
  $c_4$ for $V^A_4$ &  0.08&(0.000) & -0.01&(0.430) &  2.09&(0.021) &  1.03&(0.138) \\%
  $c_5$ for $V^A_5$ &  0.01&(0.486) & -0.03&(0.032) &  3.46&(0.000) &  1.30&(0.053) \\%
  \hline
  $d_1$ for $V^B_1$ &  {\color{red}{0.18}}&(0.000) & {\color{red}{-1.43}}&(0.000) &  {\color{red}{2.30}}&(0.005) & {\color{red}{-36.45}}&(0.000) \\%
  $d_2$ for $V^B_2$ & {\color{red}{-0.05}}&(0.049) & {\color{red}{-0.26}}&(0.000) & -1.37&(0.143) & {\color{red}{-11.77}}&(0.000) \\%
  $d_3$ for $V^B_3$ &  0.00&(0.944) & {\color{red}{-0.03}}&(0.048) &  0.66&(0.509) &  0.24&(0.751) \\%
  $d_4$ for $V^B_4$ &  0.03&(0.147) &  0.02&(0.148) & -0.60&(0.538) &  2.61&(0.000) \\%
  $d_5$ for $V^B_5$ & -0.15&(0.000) &  0.00&(0.852) &  1.80&(0.057) &  4.82&(0.000) \\%
  \hline
  $10^2e_1$ for $G^A_1$ &  {\color{red}{4.56}}&(0.000) & -0.19&(0.187) &  {\color{red}{0.51}}&(0.000) &  0.12&(0.160) \\%
  $10^2e_2$ for $G^A_2$ &  {\color{red}{1.77}}&(0.000) &  0.40&(0.099) &  {\color{red}{0.92}}&(0.002) & -0.05&(0.652) \\%
  $10^2e_3$ for $G^A_3$ &  {\color{red}{0.58}}&(0.022) &  0.77&(0.009) &  {\color{red}{-0.81}}&(0.031) &  0.24&(0.077) \\%
  $10^2e_4$ for $G^A_4$ &  0.22&(0.376) & -0.28&(0.292) &  2.37&(0.000) &  0.19&(0.216) \\%
  $10^2e_5$ for $G^A_5$ & -0.06&(0.826) &  0.51&(0.030) &  0.43&(0.429) &  0.41&(0.006) \\%
  \hline
  $10^2f_1$ for $G^B_1$ & {\color{red}{-0.73}}&(0.000) &  {\color{red}{4.50}}&(0.000) & -0.07&(0.759) &  {\color{red}{0.54}}&(0.000) \\%
  $10^2f_2$ for $G^B_2$ & {\color{red}{-1.06}}&(0.001) &  {\color{red}{2.96}}&(0.000) & -0.29&(0.346) &  {\color{red}{1.03}}&(0.000) \\%
  $10^2f_3$ for $G^B_3$ & -0.07&(0.855) & -0.29&(0.426) &  0.23&(0.595) &  {\color{red}{0.46}}&(0.024) \\%
  $10^2f_4$ for $G^B_4$ &  0.57&(0.155) & -0.01&(0.972) &  0.46&(0.304) &  {\color{red}{0.51}}&(0.015) \\%
  $10^2f_5$ for $G^B_5$ &  0.35&(0.282) & -0.48&(0.173) &  0.33&(0.416) & -0.21&(0.321) \\%
  \hline\hline
  \end{tabular}
\end{table}

\begin{figure}[tb]
\centering
\includegraphics[width=7.5cm]{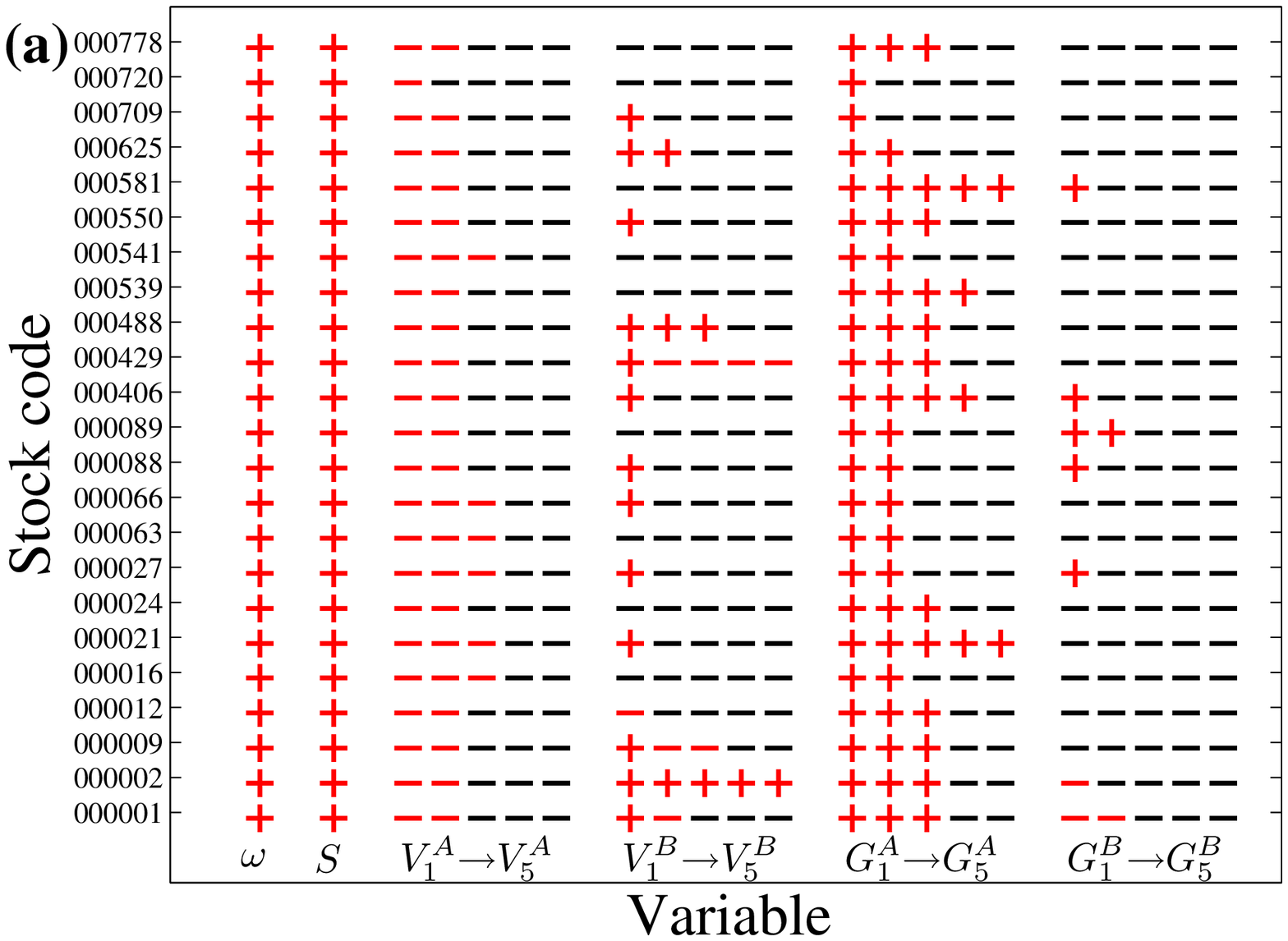}
\includegraphics[width=7.5cm]{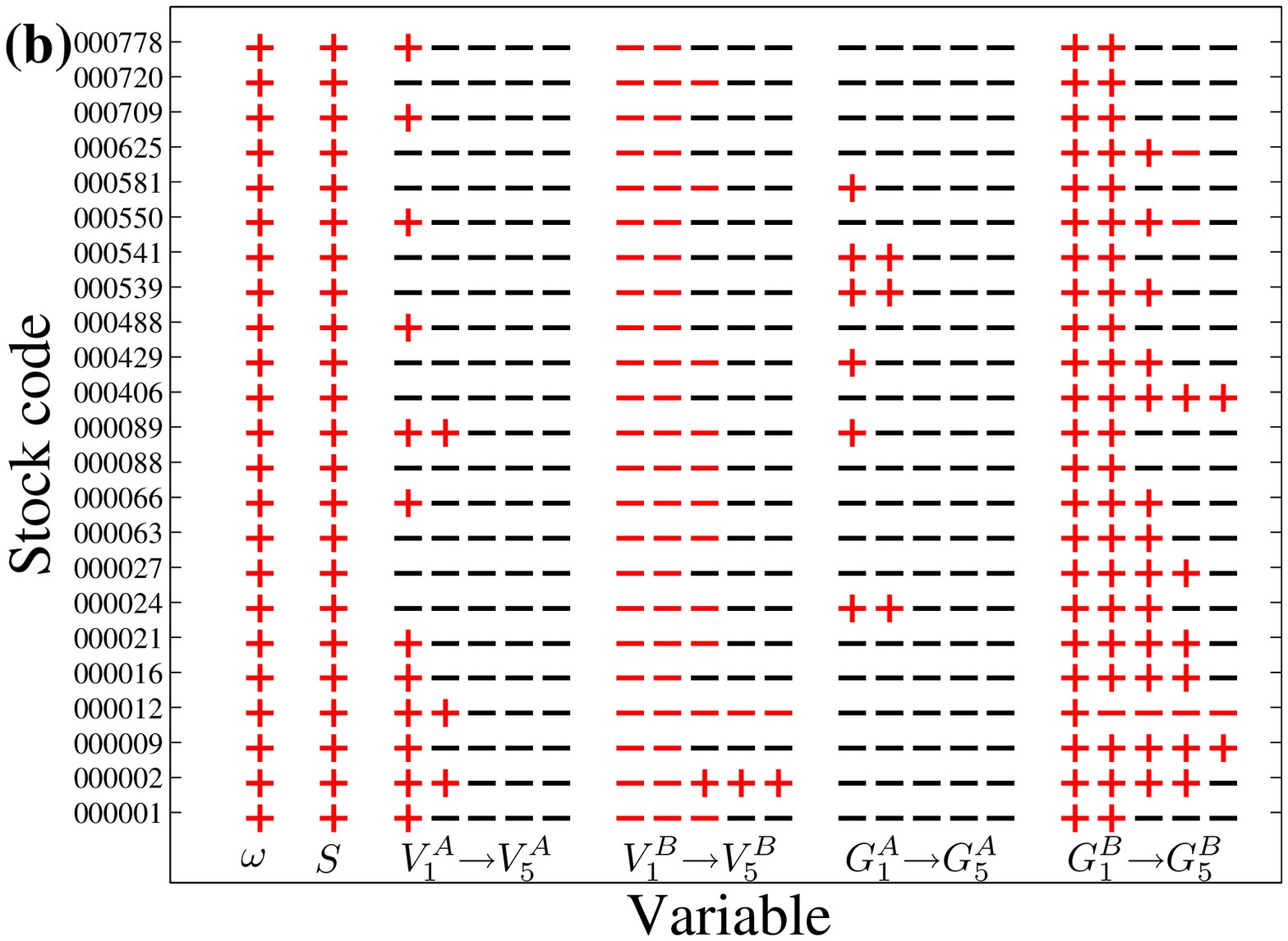}
\includegraphics[width=7.5cm]{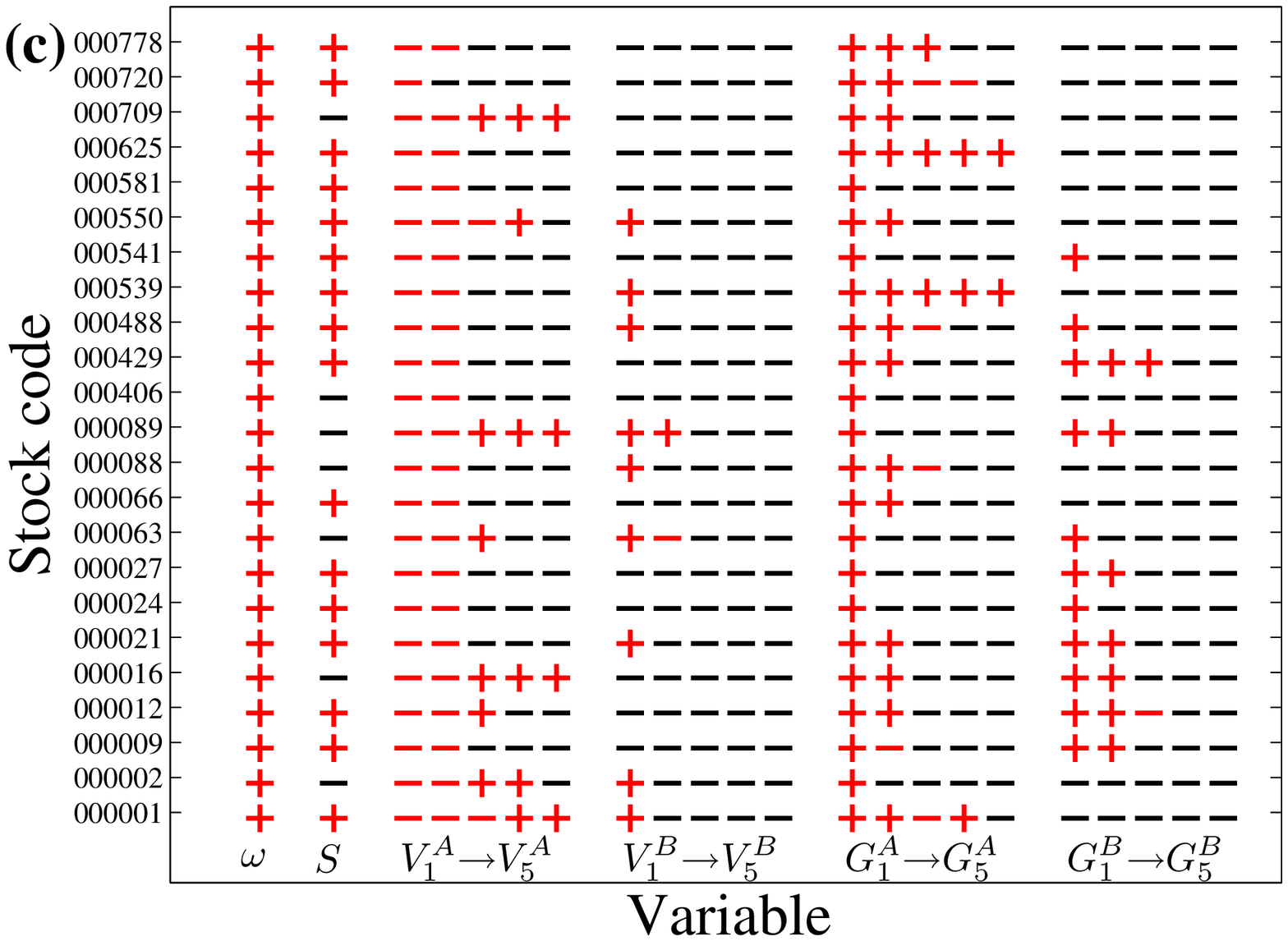}
\includegraphics[width=7.5cm]{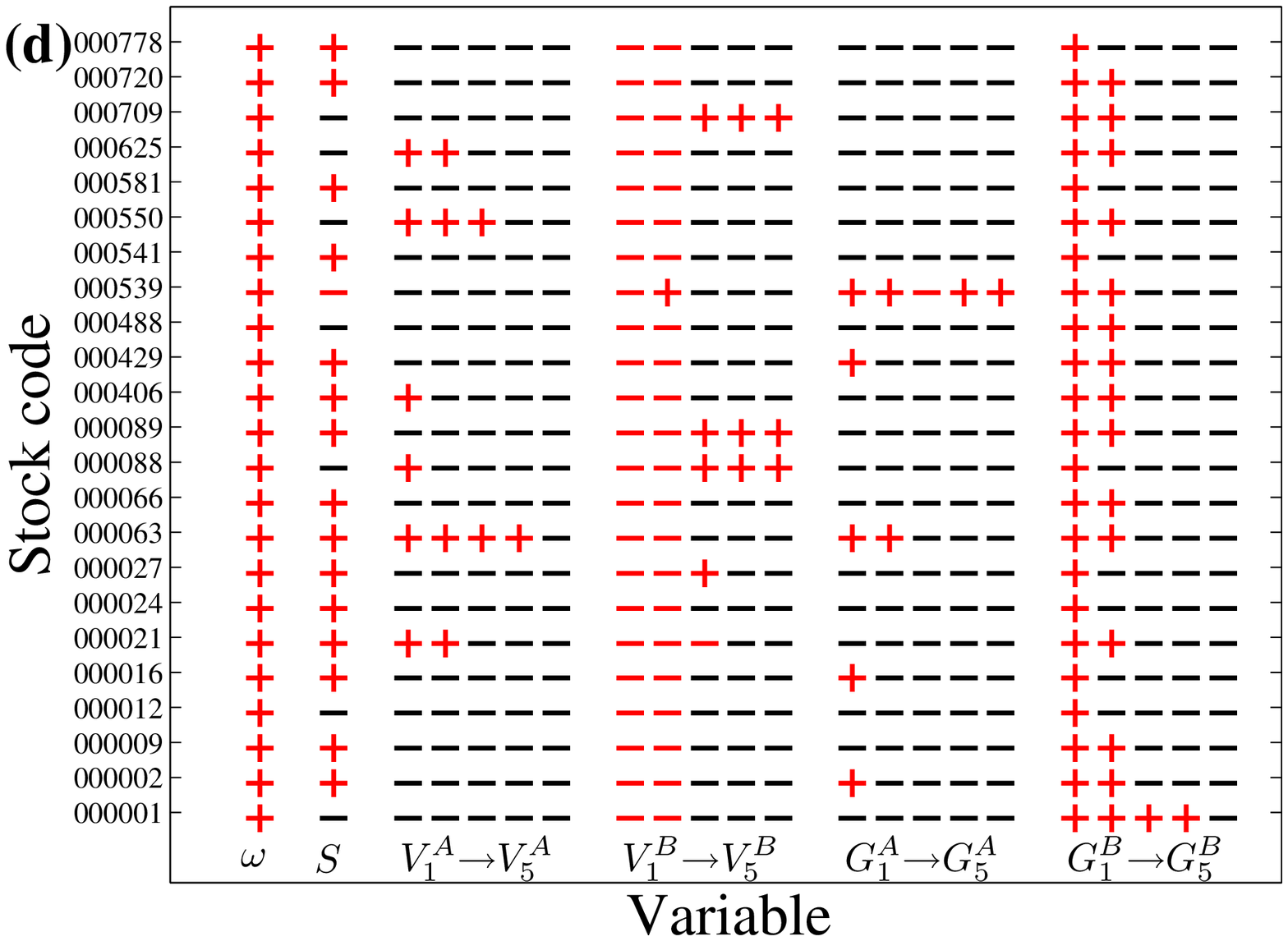}
\caption{\label{Fig:IndStocks:Level:5} Schematic illustration of coefficient significance in the power-law model (\ref{Eq:Pi:IP:Model:PL}) with $L=5$ for individual stocks: (a) buyer-initiated partially filled trades, (b) seller-initiated partially filled trades, (c) buyer-initiated filled trades, (d) seller-initiated filled trades. A plus or minus means that the corresponding coefficient is positive or negative, respectively. Red symbols ({\color{red}{$+$}} and {\color{red}{$-$}}) indicate that the corresponding variables are significant at the 5\% level, while black symbols ({\color{black}{$+$}} and {\color{black}{$-$}}) mean that the corresponding variables are insignificant at the 5\% level.}
\end{figure}

The results for other stocks are quite similar, as summarized in figure \ref{Fig:IndStocks:Level:5}, where a plus or minus symbol (+ or -) indicates that the estimated coefficient is positive or negative, and the symbol is marked in red if the associated coefficient is significant at the 5\% level. It is evident that a larger order causes a greater price impact in magnitude for all four types of trades and for all individual stocks. For partially filled trades, the price impact is larger if the bid-ask spread is larger for all stocks, while for filled trades, the spread could have a positive or negative influence on the price impact. Again, there are more coefficients on the ask side ($V_i^A$ and $G_i^A$) that are significant for buyer-initiated trades, and there are more coefficients on the bid side ($V_i^B$ and $G_i^B$) that are significant for seller-initiated trades. This finding is consistent with the mechanical analysis expressed in equations (\ref{Eq:Pi:PS}), (\ref{Eq:Pi:PB}), (\ref{Eq:Pi:FS}), and (\ref{Eq:Pi:FB}).

Note that we do not show the results of the logarithmic model for individual stocks. The dependent matrices for several stocks are is rank deficient within machine precision (singular, close to singular or badly scaled), such that no regression results can be obtained. For the stocks with nonsingular dependence matrix, the results are qualitatively similar to those obtained from the power-law model.

\subsection{The case of two levels of liquidity $L=2$}
\label{S2:IndStock:L2}

Figure \ref{Fig:IndStocks:Level:5} shows that most coefficients of $V_i^A$ and $G_i^A$ (respectively $V_i^B$ and $G_i^B$) for buyer-initiated (respectively seller-initiated) trades are insignificant when $i\geqslant 3$. We thus set $L=2$ and regress the power-law model to the data again. The results for stock 000001 are presented in table \ref{Tb:Model:PL:L2:000001}. In this case, we find that $\alpha=0.25$ and $\beta=0.15$ for buyer-initiated partially filled trades, $\alpha=0.30$ and $\beta=0.20$ for seller-initiated partially filled trades, $\alpha=0.55$ and $\beta=0.10$ for buyer-initiated filled trades, and $\alpha=0.45$ and $\beta=0.10$ for seller-initiated filled trades. The F-tests show that the model is significant for all the four types of trades and the explanatory power of the variables is very high. We observe that the values of the power-law exponents $\alpha$ and $\beta$ are closed to their counterparts in the case of $L=5$ in table \ref{Tb:Model:PL:L5:000001}.

\begin{table}[tb]
\caption{\label{Tb:Model:PL:L2:000001} Calibration of the power-law model with $L=2$ for Stock 000001. We obtain that $\alpha=0.25$ and $\beta=0.15$ for buyer-initiated partially filled trades, $\alpha=0.30$ and $\beta=0.20$ for seller-initiated partially filled trades, $\alpha=0.55$ and $\beta=0.10$ for buyer-initiated filled trades, and $\alpha=0.45$ and $\beta=0.10$ for seller-initiated filled trades. The numbers in the parentheses are the $p$-values. The estimates of the coefficients in red indicate that the associated variables are significant at the 5\% level.}
\medskip
\centering
  \begin{tabular}{cr@{ }lr@{ }lr@{ }lr@{ }l cc}
  \hline \hline
    Variable & \multicolumn{2}{c}{PB} & \multicolumn{2}{c}{PS} & \multicolumn{2}{c}{FB} & \multicolumn{2}{c}{FS} \\\hline %
    $R^2$-adj & \multicolumn{2}{c}{0.473} & \multicolumn{2}{c}{0.500} & \multicolumn{2}{c}{0.438} & \multicolumn{2}{c}{0.464} \\ %
    $p$-value & \multicolumn{2}{c}{0.000} & \multicolumn{2}{c}{0.000} & \multicolumn{2}{c}{0.000} & \multicolumn{2}{c}{0.000} \\ %
  \hline
  $a_0$ &  {\color{red}{1.08}}&(0.000) &  {\color{red}{0.44}}&(0.000) & {\color{red}{44.80}}&(0.000) & {\color{red}{19.82}}&(0.000) \\%
  $a$ for $\omega$ &  {\color{red}{1.16}}&(0.000) &  {\color{red}{1.06}}&(0.000) &  {\color{red}{1.82}}&(0.000) &  {\color{red}{2.88}}&(0.000) \\%
  $b$ for $S$ & {\color{red}{20.43}}&(0.000) & {\color{red}{31.37}}&(0.000) & {\color{red}{222.18}}&(0.000) &  3.19&(0.654) \\%
  \hline
  $c_1$ for $V^A_1$ & {\color{red}{-1.83}}&(0.000) &  {\color{red}{0.07}}&(0.000) & {\color{red}{-30.60}}&(0.000) &  {\color{red}{0.99}}&(0.000) \\%
  $c_2$ for $V^A_2$ & {\color{red}{-0.25}}&(0.000) & -0.00&(0.932) & {\color{red}{-8.52}}&(0.000) &  {\color{red}{0.96}}&(0.001) \\%
  \hline
  $d_1$ for $V^B_1$ &  {\color{red}{0.18}}&(0.000) & {\color{red}{-1.45}}&(0.000) &  {\color{red}{2.94}}&(0.000) & {\color{red}{-16.81}}&(0.000) \\%
  $d_2$ for $V^B_2$ & -0.04&(0.052) & {\color{red}{-0.26}}&(0.000) & -1.06&(0.233) & {\color{red}{-4.83}}&(0.000) \\%
  \hline
  $10^2e_1$ for $G^A_1$ &  {\color{red}{4.54}}&(0.000) & -0.17&(0.238) &  {\color{red}{0.46}}&(0.000) &  0.12&(0.182) \\%
  $10^2e_2$ for $G^A_2$ &  {\color{red}{1.82}}&(0.000) &  0.55&(0.022) &  {\color{red}{0.89}}&(0.003) & -0.04&(0.708) \\%
  \hline
  $10^2f_1$ for $G^B_1$ & {\color{red}{-0.71}}&(0.000) &  {\color{red}{4.50}}&(0.000) & -0.04&(0.853) &  {\color{red}{0.60}}&(0.000) \\%
  $10^2f_2$ for $G^B_2$ & {\color{red}{-0.89}}&(0.007) &  {\color{red}{2.95}}&(0.000) & -0.19&(0.533) &  {\color{red}{1.06}}&(0.000) \\%
  \hline\hline
  \end{tabular}
\end{table}

In table \ref{Tb:Model:PL:L2:000001}, the $G_2^A$ coefficient with the $p$-value being 0.022 for the PS trades is regarded as insignificant since the the $G_1^A$ coefficient is insignificant. Comparing table \ref{Tb:Model:PL:L2:000001} with table \ref{Tb:Model:PL:L5:000001}, we find that the corresponding coefficients are close to each other with similar significance levels. However, the coefficients of $V_1^A$ and $V_2^A$ become significant when $L$ decreases from 5 to 2. This is not surprising since the two $p$-values for $V_1^A$ and $V_2^A$ are only slightly greater than 5\% when $L=5$, and the removal of variables in the regression model changes the $p$-values slightly.

\begin{figure}[tb]
\centering
\includegraphics[width=7.5cm]{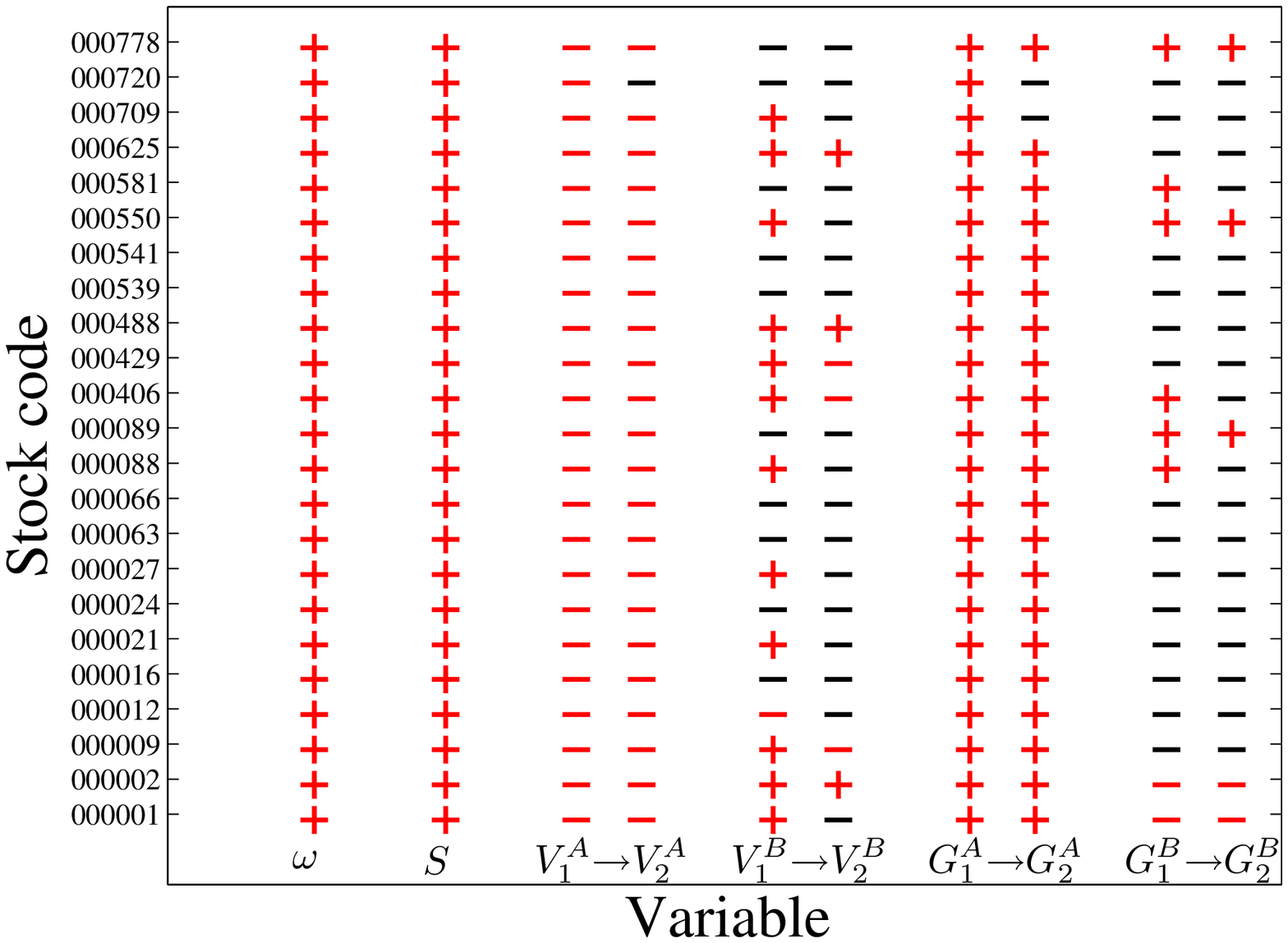}
\includegraphics[width=7.5cm]{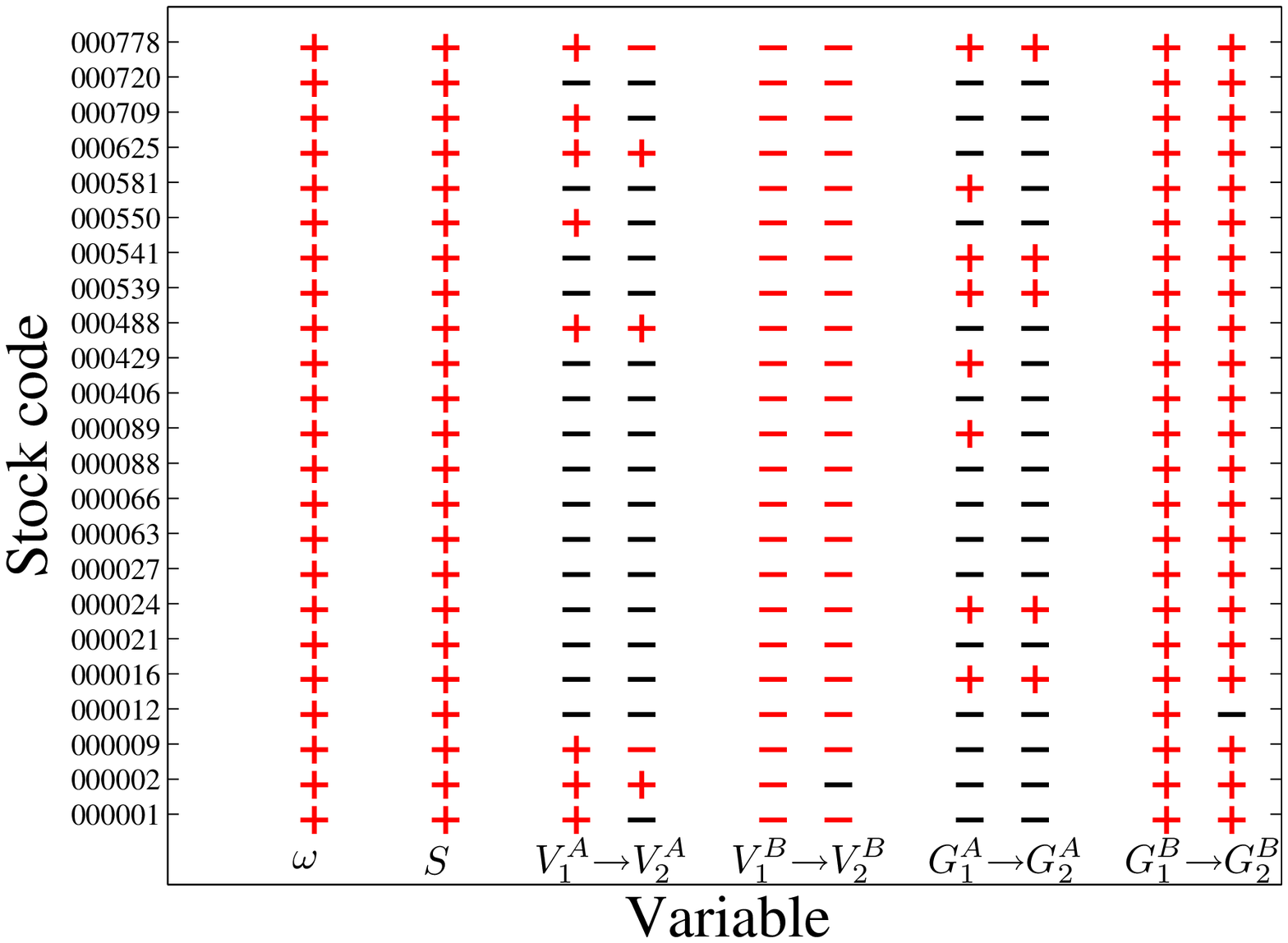}
\includegraphics[width=7.5cm]{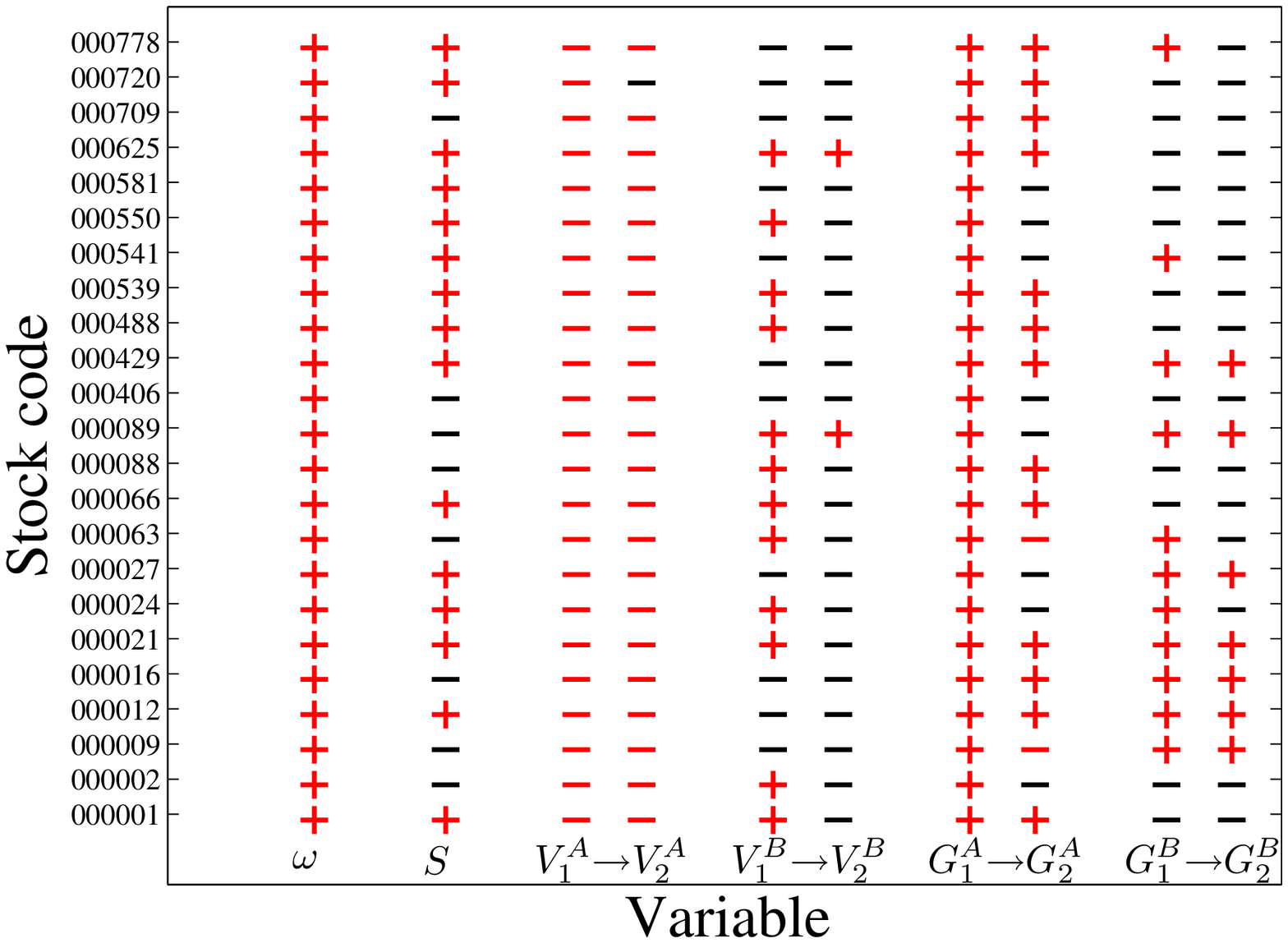}
\includegraphics[width=7.5cm]{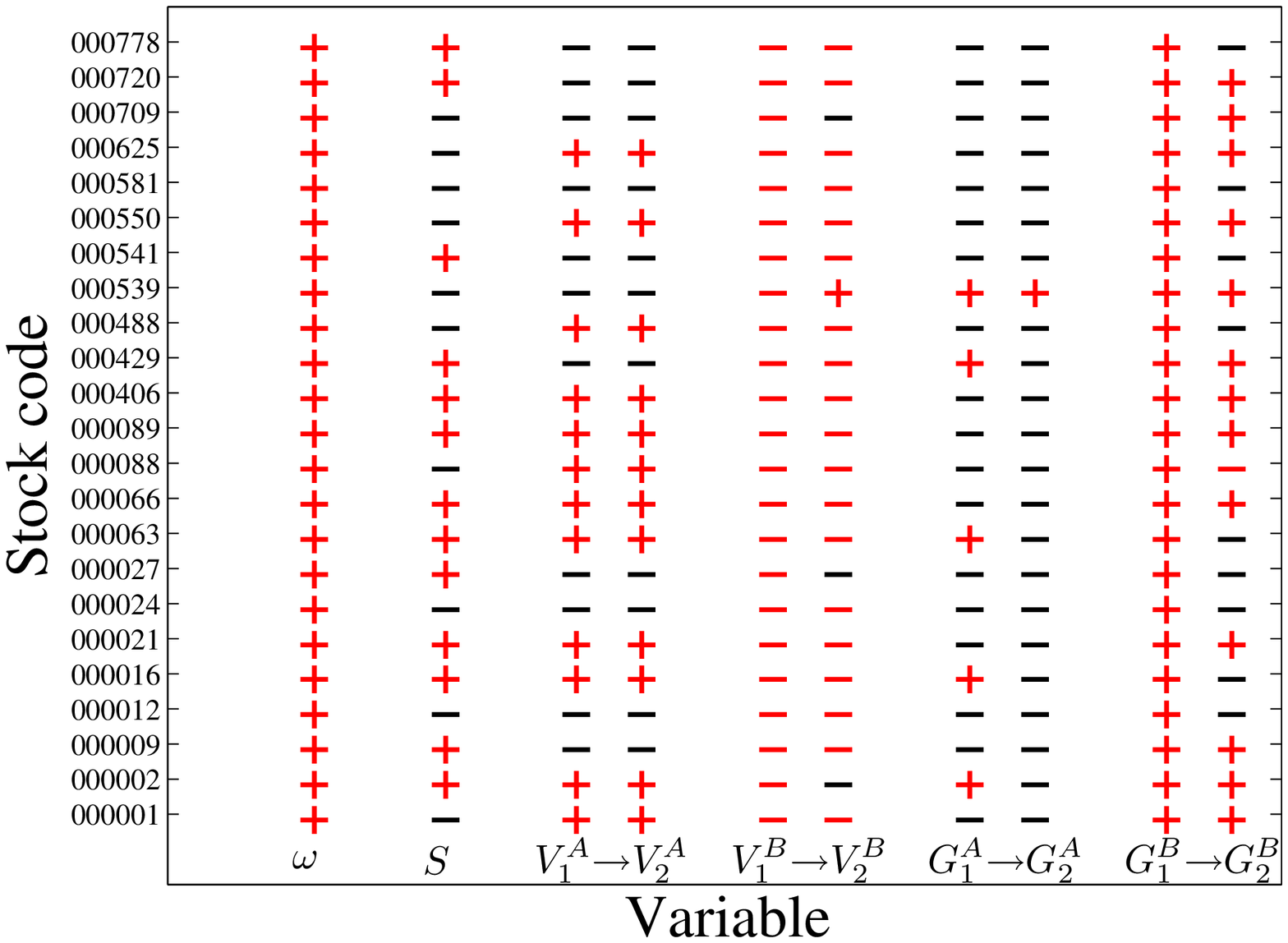}
\caption{\label{Fig:IndStocks:Level:2} Schematic illustration of coefficient significance in the power-law model (\ref{Eq:Pi:IP:Model:PL}) with $L=2$ for individual stocks: (a) buyer-initiated partially filled trades, (b) seller-initiated partially filled trades, (c) buyer-initiated filled trades, (d) seller-initiated filled trades. A plus or minus means that the corresponding coefficient is positive or negative, respectively. Red symbols ({\color{red}{$+$}} and {\color{red}{$-$}}) indicate that the corresponding variables are significant at the 5\% level, while black symbols ({\color{black}{$+$}} and {\color{black}{$-$}}) mean that the corresponding variables are insignificant at the 5\% level.}
\end{figure}

The regression results for all individual stocks are illustrated in figure \ref{Fig:IndStocks:Level:2}, where the meaning of the symbols are the same as in figure \ref{Fig:IndStocks:Level:5}. Similar patterns are observed. We can reach a conclusion that the trade size, the bid-ask spread (a liquidity measure known as market width), and the market depth at the first few price levels on the opposite limit order book plays a major role in determining the price impact of a trade.

\subsection{Asymmetrical impacts of buyer- and seller-initiated trades}
\label{S2:Asymetric}

\begin{figure}[tb]
 \centering
 \includegraphics[width=5cm]{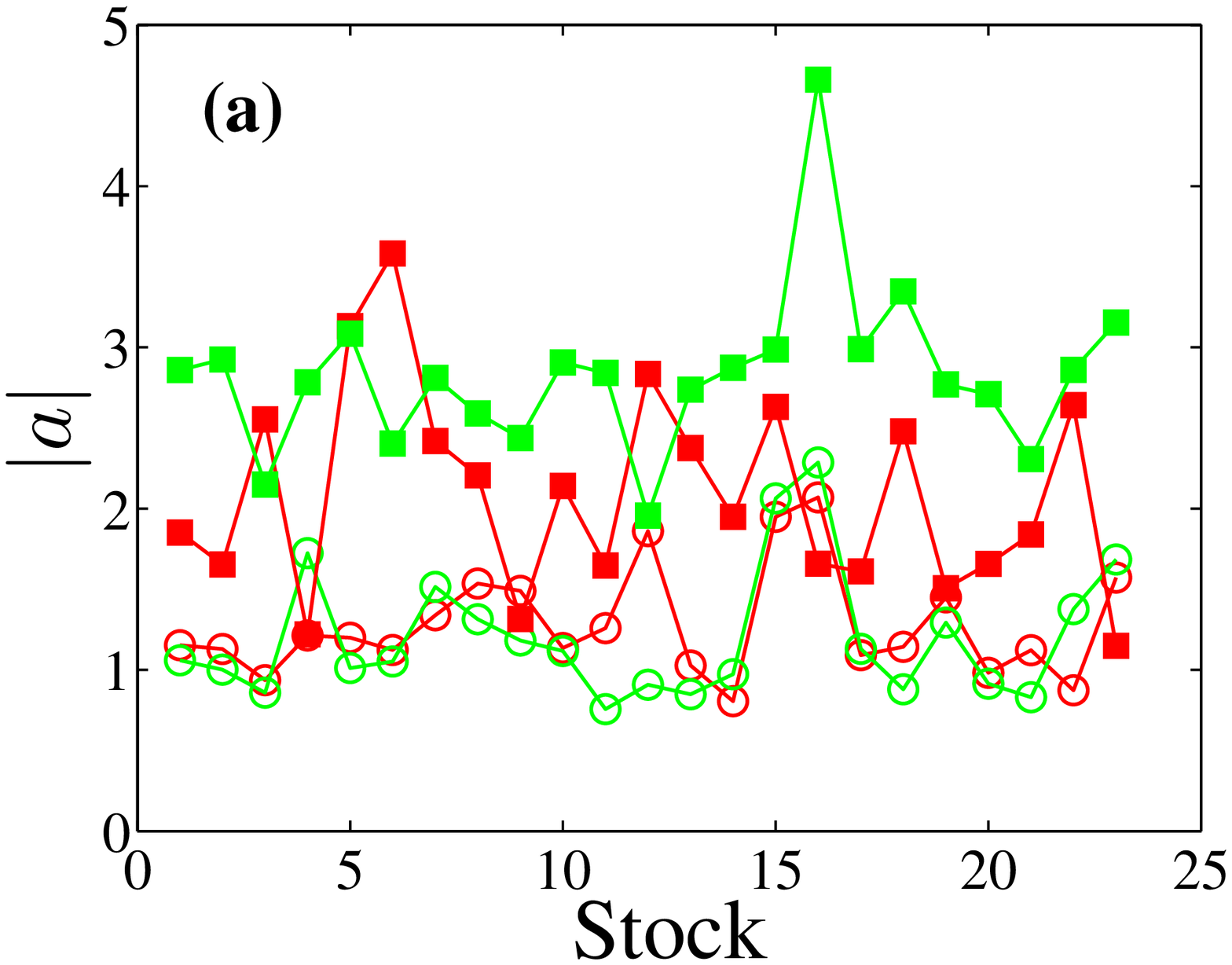}
 \includegraphics[width=5cm]{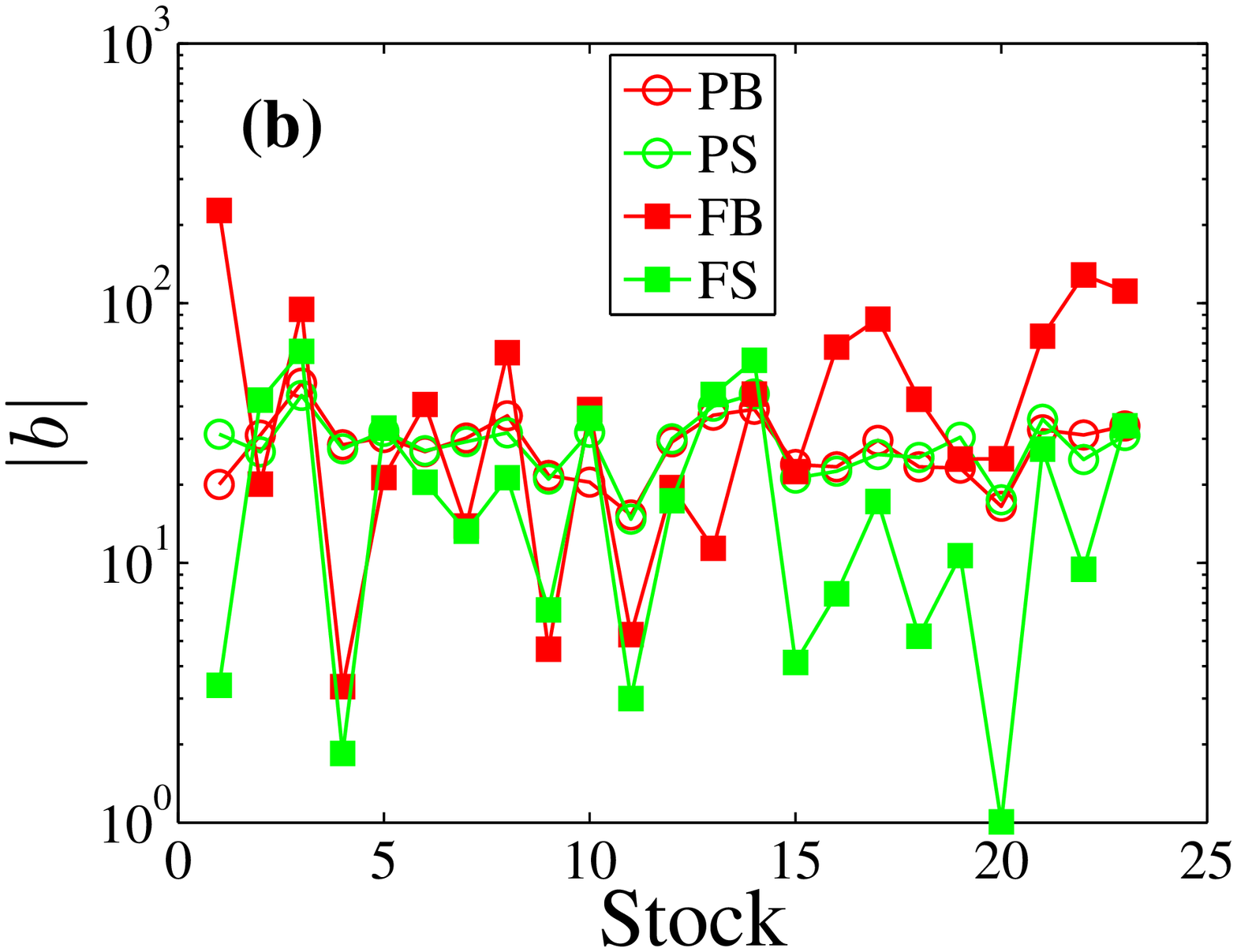}
 \includegraphics[width=5cm]{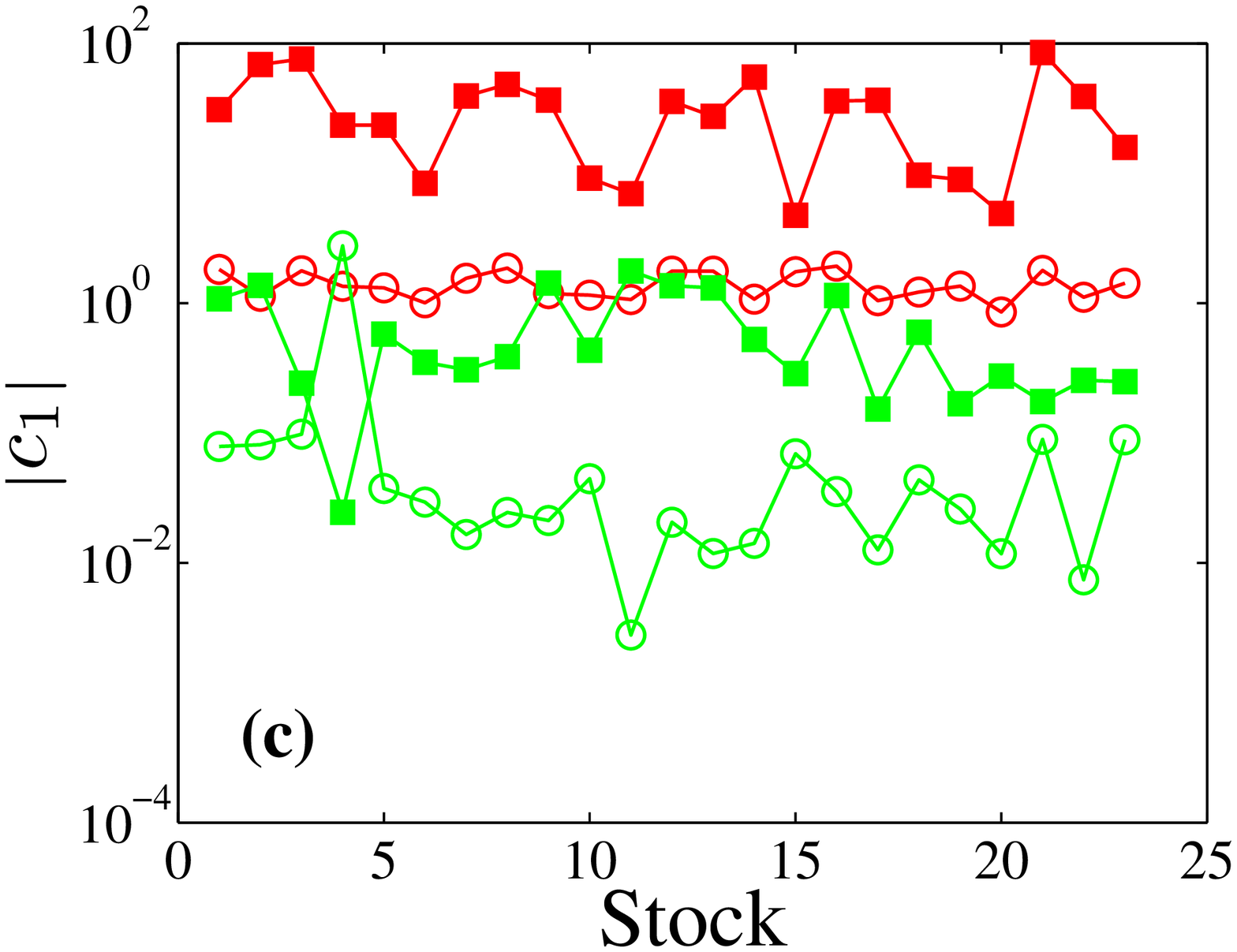}
 \includegraphics[width=5cm]{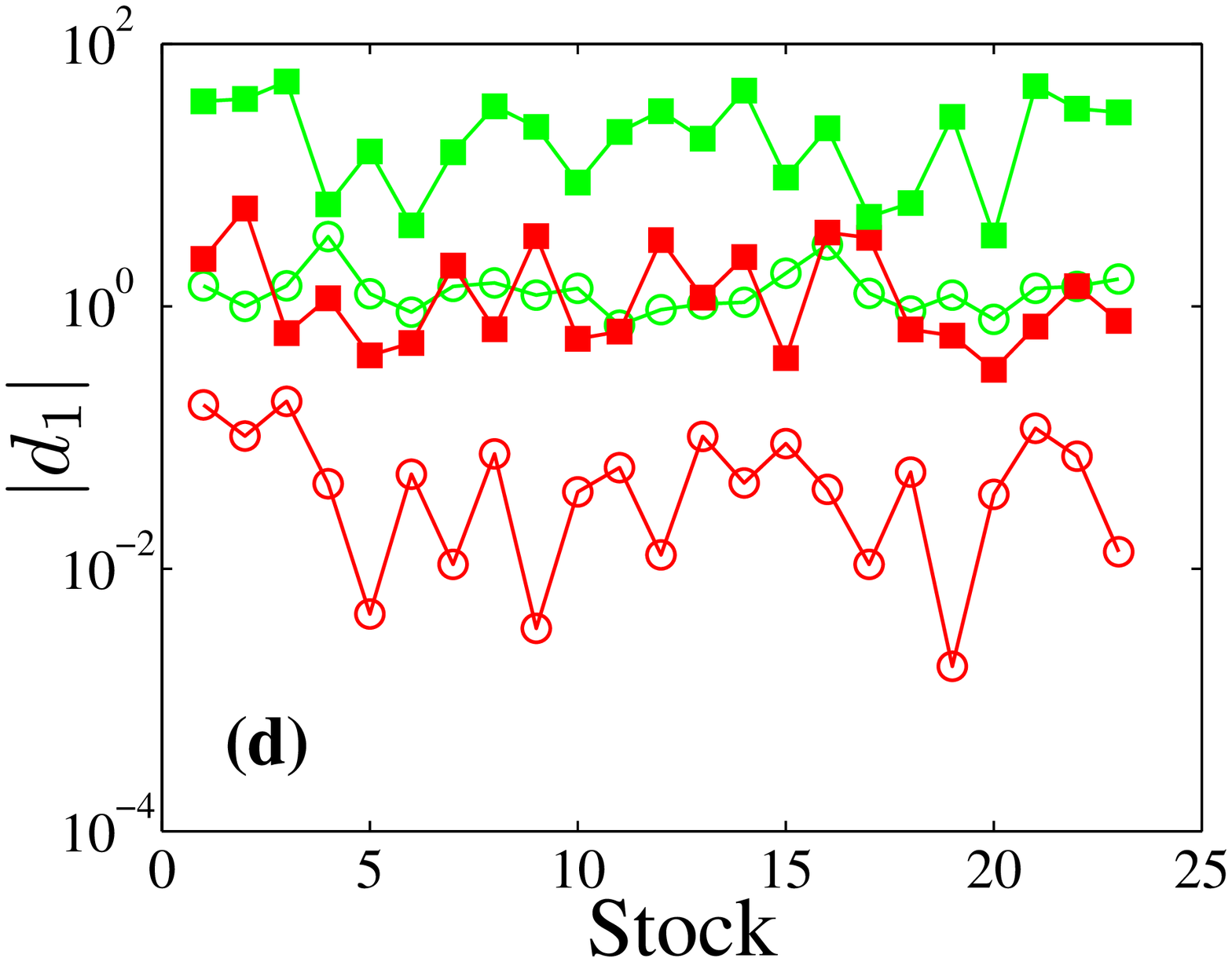}
 \includegraphics[width=5cm]{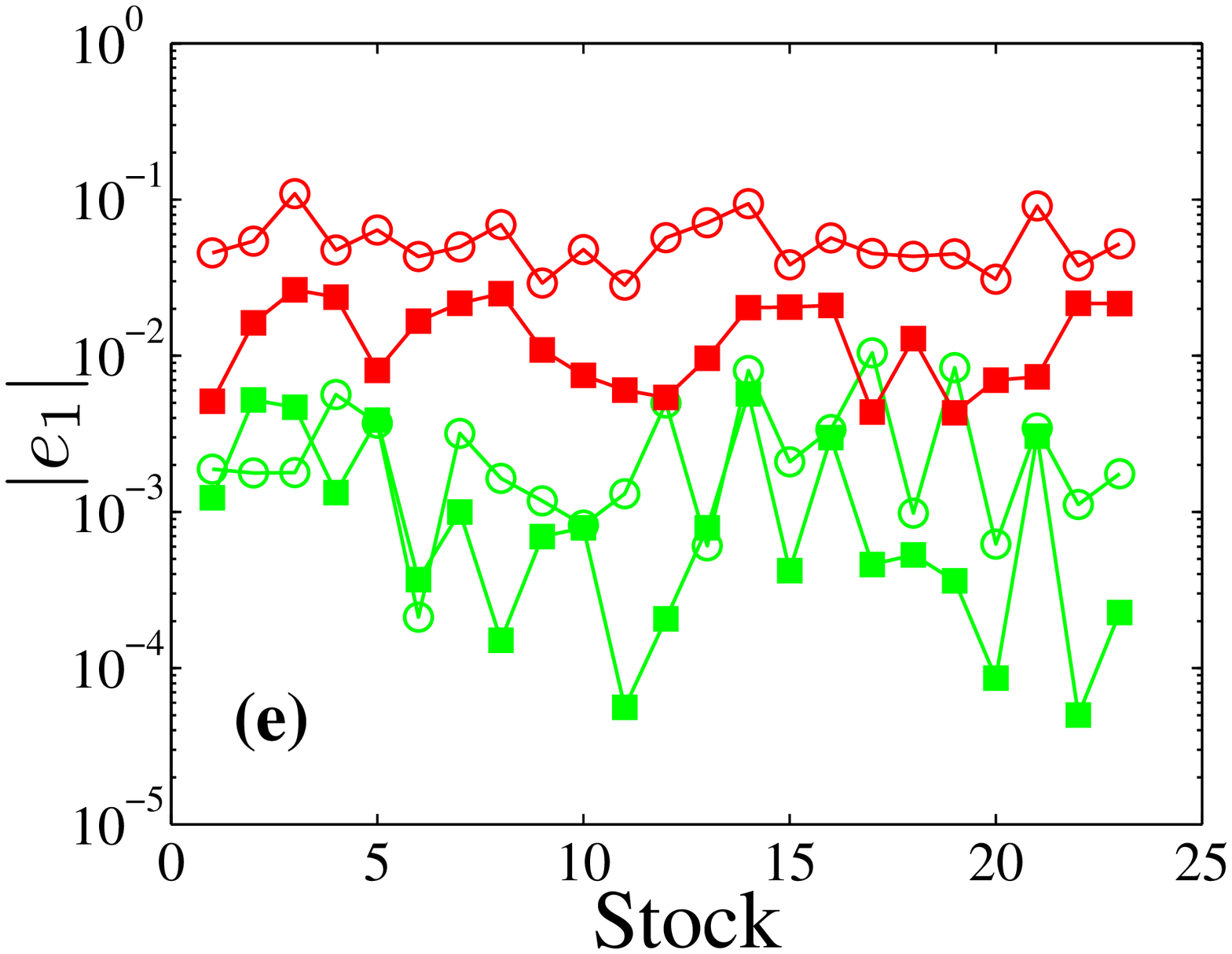}
 \includegraphics[width=5cm]{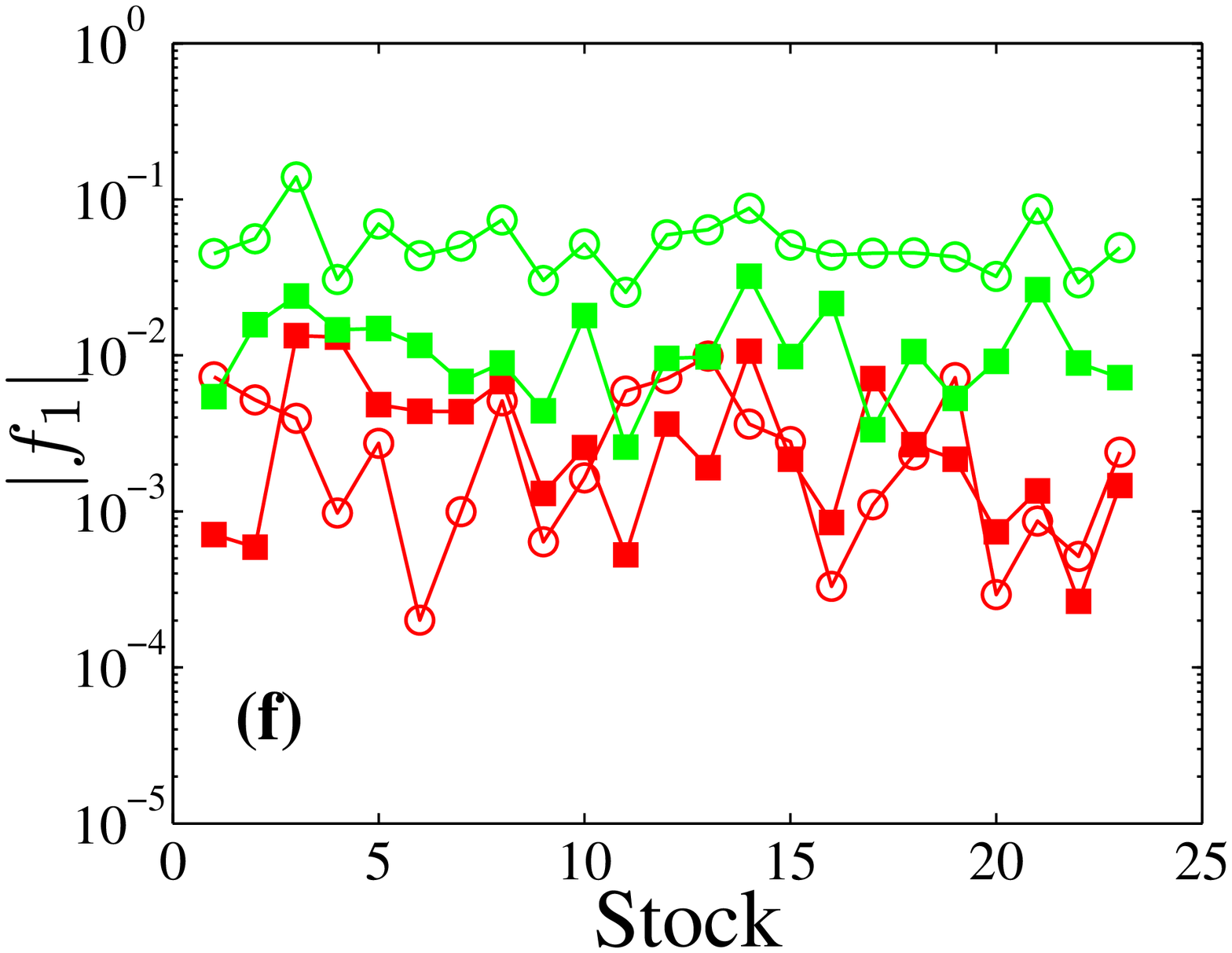}
 \caption{\label{Fig:Assymetry} Comparison of the absolute values of the coefficients ($a$, $b$, $c_1$, $d_1$, $e_1$, and $f_1$) among the four types of trades for individual stocks. Curves are red for buyer-initiated trades and green for seller-initiated trades. Symbols are open for partially filled trades and solid for filled trades.}
\end{figure}

It has long been known that market reacts differently to buy and sell orders, and sell orders have a larger price impact than purchases \cite{Saar-2001-RFS}. The regression results enable us to investigate the possible buy-sell asymmetry in the impacts of different influence factors. Figure \ref{Fig:Assymetry} shows a comparison of the absolute values of the coefficients $a$, $b$, $c_1$, $d_1$, $e_1$, and $f_1$ among the four types of trades for individual stocks.

According to figure \ref{Fig:Assymetry}(a), there is no buy-sell asymmetry caused by trade sizes for partially filled trades since the $a$ values for PB and PS trades are close to each other. In contrast, FS trades have larger $a$ values than FB trades for most stocks, such that an FS trade has a larger price impact than an FB trade of the same size.

According to figure \ref{Fig:Assymetry}(b), there is no buy-sell asymmetry caused by bid-ask spreads for partially filled trades. Concerning filled trades, the majority of the stocks exhibit buy-sell asymmetry where a lack of liquidity ({\it{i.e.}}, wide bid-ask spread) results in larger price impacts for FB trades than for FS trades.

Figure \ref{Fig:Assymetry}(c,e) shows that the depth of the limit order book on the ask side, $V_i^A$ and $G_i^A$, has larger price impacts on buyer-initiated trades than on seller-initiated trades. In contrast, figure \ref{Fig:Assymetry}(d,f) shows that the depth of the limit order book on the bid side, $V_i^B$ and $G_i^B$, has larger price impacts on seller-initiated trades than on buyer-initiated trades. These observations are consistent with the results obtained in previous sections, stating that the liquidity at the opposite side has greater influence on the price impact than the liquidity at the same side.

\section{Cross-sectional regression for all stocks}
\label{S1:AllStocks}

In section \ref{S1:IndStocks}, we have performed analysis on individual stocks. In this section, we perform a cross-sectional analysis of all stocks. Both the power-law model and the logarithmic model are considered. In the regression, the normalized data for each variable are put together correspondingly and we use $L=5$ in the two models.

\subsection{The power-law model}

Regressing the power-law model, we obtain that $\alpha=0.25$ and $\beta=0.20$ for buyer-initiated partially filled trades, $\alpha=0.30$ and $\beta=0.20$ for seller-initiated partially filled trades, $\alpha=0.55$ and $\beta=0.05$ for buyer-initiated filled trades, and $\alpha=0.50$ and $\beta=0.05$ for seller-initiated filled trades. The estimated coefficients are presented in table \ref{Tb:Model:PL:L5}, where the estimates that are significant at the 5\% level are marked in red. The R-squares from the F-tests show that the model is significant for all four types of trades. We observe that the determinants ($\omega$, $S$, $V_i^A$, $V_i^B$, $G_i^A$ and $G_i^B$) and the dummy variables can account for 43.6\% of the price impact for PB trades, 22.2\% for PS trades, 40.2\% for FB trades, and 39.6\% for FS trades.

\begin{table}[tb]
\caption{\label{Tb:Model:PL:L5} Cross-sectional regression results of the power-law model in (\ref{Eq:Pi:IP:Model:PL}) with $L=5$ for all stocks. We obtain that $\alpha=0.25$ and $\beta=0.20$ for buyer-initiated partially filled trades, $\alpha=0.30$ and $\beta=0.20$ for seller-initiated partially filled trades, $\alpha=0.55$ and $\beta=0.05$ for buyer-initiated filled trades, and $\alpha=0.50$ and $\beta=0.05$ for seller-initiated filled trades. The numbers in the parentheses are the $p$-values. The estimates of the coefficients in red indicate that the associated variables are significant at the 5\% level.}
\medskip
\centering
  \begin{tabular}{cr@{ }lr@{ }lr@{ }lr@{ }l cc}
  \hline \hline
    Variable & \multicolumn{2}{c}{PB} & \multicolumn{2}{c}{PS} & \multicolumn{2}{c}{FB} & \multicolumn{2}{c}{FS} \\\hline %
    $R^2$-adj & \multicolumn{2}{c}{0.436} & \multicolumn{2}{c}{0.222} & \multicolumn{2}{c}{0.402} & \multicolumn{2}{c}{0.396} \\ %
    $p$-value & \multicolumn{2}{c}{0.000} & \multicolumn{2}{c}{0.000} & \multicolumn{2}{c}{0.000} & \multicolumn{2}{c}{0.000} \\ %
  \hline
  $a_0$ &  {\color{red}0.40}&(0.000) &  {\color{red}0.90}&(0.000) & {\color{red}34.74}&(0.000) & {\color{red}18.67}&(0.000) \\%
  $a$ for $\omega$ &  {\color{red}1.26}&(0.000) &  {\color{red}1.15}&(0.000) &  {\color{red}1.81}&(0.000) &  {\color{red}2.49}&(0.000) \\%
  $b$ for $S$ & {\color{red}20.48}&(0.000) & {\color{red}20.96}&(0.000) &  {\color{red}6.45}&(0.000) & -1.39&(0.149) \\%
  \hline
  $c_1$ for $V^A_1$ & {\color{red}-1.423}&(0.000) & {\color{red}0.084}&(0.000) & {\color{red}-48.380}&(0.000) & {\color{red}1.242}&(0.000) \\%
  $c_2$ for $V^A_2$ & {\color{red}-0.139}&(0.000) & {\color{red}0.062}&(0.000) & {\color{red}-10.806}&(0.000) & {\color{red}1.623}&(0.000) \\%
  $c_3$ for $V^A_3$ & {\color{red}0.027}&(0.000) & 0.009&(0.364) & {\color{red}3.947}&(0.000) & {\color{red}1.421}&(0.000) \\%
  $c_4$ for $V^A_4$ & {\color{red}0.048}&(0.000) & 0.028&(0.004) & {\color{red}5.312}&(0.000) & {\color{red}1.150}&(0.000) \\%
  $c_5$ for $V^A_5$ & {\color{red}0.024}&(0.000) & -0.010&(0.273) & {\color{red}4.390}&(0.000) & {\color{red}1.454}&(0.000) \\%
  \hline
  $d_1$ for $V^B_1$ & {\color{red}0.099}&(0.000) & {\color{red}-1.532}&(0.000) & {\color{red}3.273}&(0.000) & {\color{red}-31.612}&(0.000) \\%
  $d_2$ for $V^B_2$ & {\color{red}0.062}&(0.000) & {\color{red}-0.194}&(0.000) & {\color{red}1.645}&(0.000) & {\color{red}-5.179}&(0.000) \\%
  $d_3$ for $V^B_3$ & {\color{red}0.047}&(0.000) & -0.000&(0.984) & {\color{red}1.978}&(0.000) & {\color{red}2.195}&(0.000) \\%
  $d_4$ for $V^B_4$ & {\color{red}0.042}&(0.000) & 0.031&(0.001) & {\color{red}1.453}&(0.000) & {\color{red}3.703}&(0.000) \\%
  $d_5$ for $V^B_5$ & {\color{red}0.022}&(0.000) & 0.015&(0.111) & {\color{red}3.123}&(0.000) & {\color{red}4.052}&(0.000) \\%
  \hline
  $10^2e_1$ for $G^A_1$ & {\color{red}3.785}&(0.000) & {\color{red}-0.360}&(0.000) & {\color{red}0.957}&(0.000) & {\color{red}0.053}&(0.000) \\%
  $10^2e_2$ for $G^A_2$ & {\color{red}0.569}&(0.000) & {\color{red}-0.367}&(0.000) & {\color{red}0.361}&(0.000) & {\color{red}0.041}&(0.012) \\%
  $10^2e_3$ for $G^A_3$ & {\color{red}-0.195}&(0.000) & {\color{red}-0.411}&(0.000) & {\color{red}0.361}&(0.000) & {\color{red}0.038}&(0.036) \\%
  $10^2e_4$ for $G^A_4$ & {\color{red}-0.431}&(0.000) & {\color{red}-0.354}&(0.000) & -0.055&(0.259) & 0.004&(0.822) \\%
  $10^2e_5$ for $G^A_5$ & {\color{red}-0.411}&(0.000) & {\color{red}-0.391}&(0.000) & 0.030&(0.531) & 0.042&(0.016) \\%
  \hline
  $10^2f_1$ for $G^B_1$ & {\color{red}-0.239}&(0.000) & {\color{red}3.475}&(0.000) & {\color{red}0.142}&(0.000) & {\color{red}0.721}&(0.000) \\%
  $10^2f_2$ for $G^B_2$ & {\color{red}-0.362}&(0.000) & {\color{red}0.508}&(0.000) & 0.052&(0.125) & {\color{red}0.217}&(0.000) \\%
  $10^2f_3$ for $G^B_3$ & {\color{red}-0.487}&(0.000) & {\color{red}-0.263}&(0.000) & -0.077&(0.040) & 0.028&(0.156) \\%
  $10^2f_4$ for $G^B_4$ & {\color{red}-0.249}&(0.000) & {\color{red}-0.551}&(0.000) & -0.021&(0.583) & -0.037&(0.062) \\%
  $10^2f_5$ for $G^B_5$ & {\color{red}-0.386}&(0.000) & {\color{red}-0.413}&(0.000) & -0.024&(0.501) & -0.042&(0.008) \\%
  \hline\hline
  \end{tabular}
\end{table}

For buyer-initiated partially filled trades, the normalized price impact is positively correlated with the trade size $\omega$, the bid-ask spread $S$ and the limit order volumes $V_i^B$ at the bid side, and negatively correlated with the price gaps $G_i^B$ on the bid limit order book. For the two liquidity measures at the ask side, the situation is a little bit complicated. We observe that $c_1$ and $c_2$ are negative and $c_3$, $c_4$ and $c_5$ are positive for the ask limit order volumes, and  $e_1$ and $e_2$ are positive and $e_3$, $e_4$ and $e_5$ are negative for the ask price gaps. However, it is found that $|c_1+c_2|\gg |c_3+c_4+c_5|$ and $|e_1+e_2|\gg |e_3+e_4+e_5|$, which means that the first two levels dominate. We thus argue that the normalized price impact is negatively correlated with the ask order volume and positively correlated with the ask price gaps. Further more, table \ref{Tb:Model:PL:L5} shows that $|c_1+c_2+c_3+c_4+c_5|\gg|d_1+d_2+d_3+d_4+d_5|$ and $|e_1+e_2+e_3+e_4+e_5|\gg|f_1+f_2+f_3+f_4+f_5|$, which implies that the liquidity at the ask side of the limit order book plays a major role on the price impact when compared with the liquidity at the bid side.

Similarly, for seller-initiated partially filled trades, the normalized price impact is positively correlated with the trade size $\omega$, the bid-ask spread $S$, the ask volumes and the bid price gaps on the book, and is negatively correlated with the bid volumes and the ask price gaps on the limit order book. In addition, the liquidity at the opposite side of the order book plays a more influencing role than the liquidity at the same side.

For filled trades, large trade size, wide bid-ask spread, large market depth at the same order book side, and small market depth at the opposite order book side may cause a larger normalized price impact. There are three exceptions. First, the coefficient $f_1$ for FB trades is positive, which is nevertheless much smaller than $e_1+e_2+e_3$. Second, the coefficients $e_1$, $e_2$ and $e_3$ for FS trades are positive, whose sum is however much smaller than $f_1+f_2$. Third, the coefficient $b$ for FS trades is insignificant.

\subsection{The logarithmic model}

We also perform cross-sectional regression of the logarithmic model expressed in equation (\ref{Eq:Pi:IP:Model:LN}). We obtain that $\alpha=0.15$ for buyer-initiated partially filled trades, $\alpha=0.20$ for seller-initiated partially filled trades, $\alpha=0.55$ for buyer-initiated filled trades, and $\alpha=0.50$ for seller-initiated filled trades. The results are presented in table \ref{Tb:Model:LN:L5}. The $p$-values of the F-tests indicate that the model is significant for all four types of trades. The determinants ($\omega$, $S$, $V_i^A$, $V_i^B$, $G_i^A$ and $G_i^B$) and the dummy variables can account for 42.5\% of the price impact for PB trades, 21.6\% for PS trades, 40.1\% for FB trades, and 39.6\% for FS trades.

\begin{table}[tb]
\caption{\label{Tb:Model:LN:L5} Cross-sectional regression results of the logarithmic model in (\ref{Eq:Pi:IP:Model:LN}) with $L=5$ for all stocks. We obtain that $\alpha=0.15$ for buyer-initiated partially filled trades, $\alpha=0.20$ for seller-initiated partially filled trades, $\alpha=0.55$ for buyer-initiated filled trades, and $\alpha=0.50$ for seller-initiated filled trades. The numbers in the parentheses are the $p$-values. The estimates of the coefficients in red indicate that the associated variables are significant at the 5\% level.}
\medskip
\centering
  \begin{tabular}{cr@{ }lr@{ }lr@{ }lr@{ }l cc}
  \hline \hline
    Variable & \multicolumn{2}{c}{PB} & \multicolumn{2}{c}{PS} & \multicolumn{2}{c}{FB} & \multicolumn{2}{c}{FS} \\\hline %
    $R^2$-adj & \multicolumn{2}{c}{0.425} & \multicolumn{2}{c}{0.216} & \multicolumn{2}{c}{0.401} & \multicolumn{2}{c}{0.396} \\ %
    $p$-value & \multicolumn{2}{c}{0.000} & \multicolumn{2}{c}{0.000} & \multicolumn{2}{c}{0.000} & \multicolumn{2}{c}{0.000} \\ %
  \hline
  $a_0$ & {\color{red}-1.60}&(0.000) & {\color{red}-1.16}&(0.000) &  {\color{red}0.80}&(0.000) & {\color{red}-1.28}&(0.000) \\%
  $a$ for $\omega$ &  {\color{red}2.06}&(0.000) &  {\color{red}1.70}&(0.000) &  {\color{red}1.78}&(0.000) &  {\color{red}2.48}&(0.000) \\%
  $b$ for $S$ & {\color{red}20.55}&(0.000) & {\color{red}20.89}&(0.000) &  {\color{red}6.42}&(0.000) & -1.58&(0.101) \\%
  \hline
  $c_1$ for $V^A_1$ & {\color{red}-0.254}&(0.000) & {\color{red}0.022}&(0.000) & {\color{red}-2.617}&(0.000) & {\color{red}0.064}&(0.000) \\%
  $c_2$ for $V^A_2$ & {\color{red}-0.031}&(0.000) & {\color{red}0.015}&(0.000) & {\color{red}-0.595}&(0.000) & {\color{red}0.092}&(0.000) \\%
  $c_3$ for $V^A_3$ & {\color{red} 0.005}&(0.000) &  0.003&(0.245) & {\color{red}0.204}&(0.000) & {\color{red}0.080}&(0.000) \\%
  $c_4$ for $V^A_4$ & {\color{red} 0.011}&(0.000) &  0.006&(0.004) & {\color{red}0.302}&(0.000) & {\color{red}0.067}&(0.000) \\%
  $c_5$ for $V^A_5$ & {\color{red} 0.005}&(0.000) & -0.001&(0.661) & {\color{red}0.260}&(0.000) & {\color{red}0.080}&(0.000) \\%
  \hline
  $d_1$ for $V^B_1$ & {\color{red}0.024}&(0.000) & {\color{red}-0.266}&(0.000) & {\color{red}0.173}&(0.000) & {\color{red}-1.676}&(0.000) \\%
  $d_2$ for $V^B_2$ & {\color{red}0.014}&(0.000) & {\color{red}-0.047}&(0.000) & {\color{red}0.079}&(0.000) & {\color{red}-0.279}&(0.000) \\%
  $d_3$ for $V^B_3$ & {\color{red}0.011}&(0.000) & {\color{red}-0.005}&(0.035) & {\color{red}0.104}&(0.000) & {\color{red}0.110}&(0.000) \\%
  $d_4$ for $V^B_4$ & {\color{red}0.009}&(0.000) &  0.004&(0.077) & {\color{red}0.074}&(0.000) & {\color{red}0.196}&(0.000) \\%
  $d_5$ for $V^B_5$ & {\color{red}0.005}&(0.000) & -0.001&(0.783) & {\color{red}0.164}&(0.000) & {\color{red}0.213}&(0.000) \\%
  \hline
  $10^2e_1$ for $G^A_1$ & {\color{red} 3.775}&(0.000) & {\color{red}-0.352}&(0.000) & {\color{red}0.936}&(0.000) & {\color{red}0.050}&(0.000) \\%
  $10^2e_2$ for $G^A_2$ & {\color{red} 0.552}&(0.000) & {\color{red}-0.349}&(0.000) & {\color{red}0.352}&(0.000) & {\color{red}0.042}&(0.009) \\%
  $10^2e_3$ for $G^A_3$ & {\color{red}-0.188}&(0.000) & {\color{red}-0.394}&(0.000) & {\color{red}0.371}&(0.000) & {\color{red}0.038}&(0.036) \\%
  $10^2e_4$ for $G^A_4$ & {\color{red}-0.434}&(0.000) & {\color{red}-0.331}&(0.000) & -0.043&(0.376) & 0.004&(0.809) \\%
  $10^2e_4$ for $G^A_5$ & {\color{red}-0.405}&(0.000) & {\color{red}-0.386}&(0.000) & 0.039&(0.417) & 0.042&(0.017) \\%
  \hline
  $10^2f_1$ for $G^B_1$ & {\color{red}-0.231}&(0.000) & {\color{red}3.448}&(0.000) & {\color{red}0.136}&(0.000) & {\color{red}0.716}&(0.000) \\%
  $10^2f_2$ for $G^B_2$ & {\color{red}-0.348}&(0.000) & {\color{red}0.485}&(0.000) & 0.052&(0.128) & {\color{red}0.211}&(0.000) \\%
  $10^2f_3$ for $G^B_3$ & {\color{red}-0.479}&(0.000) & {\color{red}-0.296}&(0.000) & -0.077&(0.040) & 0.024&(0.230) \\%
  $10^2f_4$ for $G^B_4$ & {\color{red}-0.241}&(0.000) & {\color{red}-0.564}&(0.000) & -0.027&(0.481) & -0.036&(0.069) \\%
  $10^2f_5$ for $G^B_5$ & {\color{red}-0.393}&(0.000) & {\color{red}-0.407}&(0.000) & -0.026&(0.462) & -0.039&(0.012) \\%
  \hline\hline
  \end{tabular}
\end{table}

Comparing the estimated coefficients (excluding $a_0$) in table \ref{Tb:Model:LN:L5} with those in table \ref{Tb:Model:PL:L5}, we find that they have exactly the same pattern, except that $b_3$ for PS trades becomes significant in the logarithmic model. Therefore, the two models lead to the same conclusions.

\subsection{The connection between the two models}

\begin{figure}[tb]
 \centering
 \includegraphics[width=7.3cm]{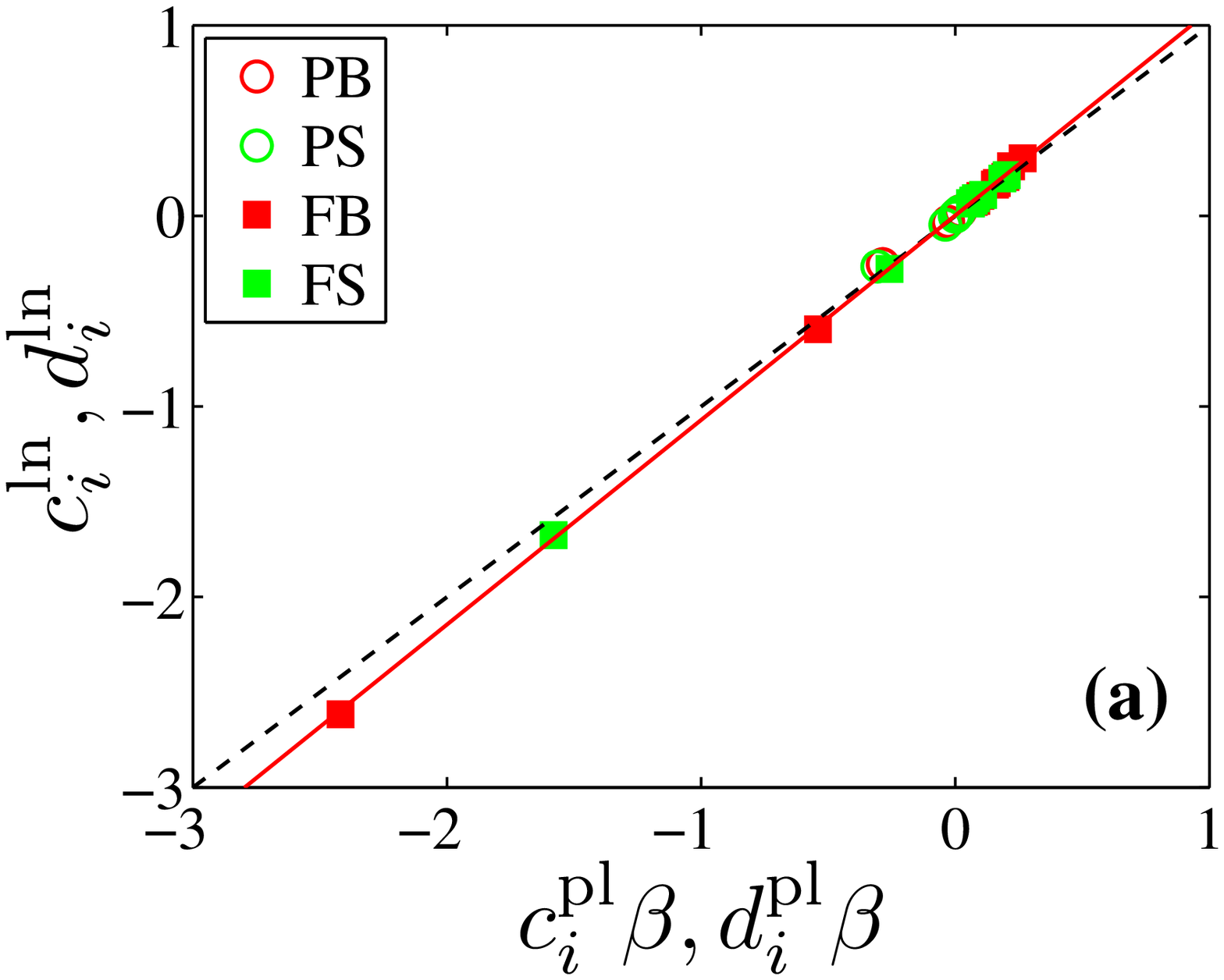}
 \includegraphics[width=7.5cm]{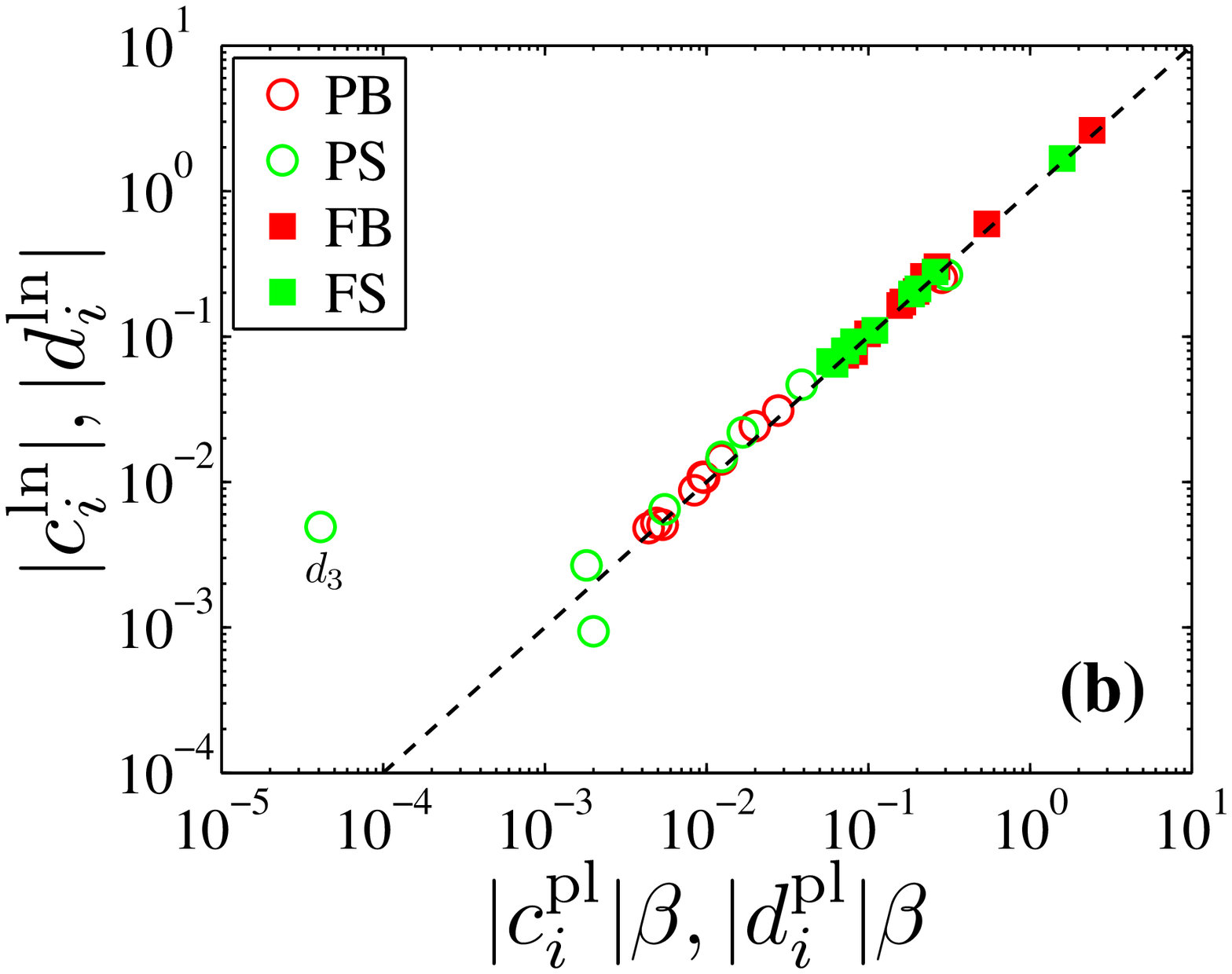}
 \caption{\label{Fig:AllStock:PL:LN} Scatter plot of $c_i^{\rm{ln}}$ against $c_i^{\rm{pl}}\beta$ and $d_i^{\rm{ln}}$ against $d_i^{\rm{pl}}\beta$. The dashed lines are $y=x$ and the solid line is a linear fit.}
\end{figure}

The difference between the two models is that there are power-law terms of the volumes in the power-law model (\ref{Eq:Pi:IP:Model:PL}) and alternative logarithmic terms of the volumes in the logarithmic model (\ref{Eq:Pi:IP:Model:LN}). If $0<\beta\ll 1$, we have
\begin{equation}
 V^\beta=e^{\beta\ln{V}}\approx 1+\beta\ln{V}.
\end{equation}
It follows that equation (\ref{Eq:Pi:IP:Model:PL}) becomes
\begin{eqnarray}
\frac{r_{k,t+1}}{\langle{r_k}\rangle}
          &=&
              a_0 + \sum_{i=1}^{L} (c_i+d_i) + a \left(\frac{\omega_{k,t}}{\langle{\omega_k}\rangle}\right)^\alpha  + b S_{k,t}   %
           + \sum_{i=1}^{L} c_i\beta \ln\left(\frac{V^A_{k,i,t}}{\langle{\omega_k}\rangle}\right) %
           \nonumber\\&&
           + \sum_{i=1}^{L} d_i\beta \ln\left(\frac{V^B_{k,i,t}}{\langle{\omega_k}\rangle}\right) %
           + \sum_{i=1}^{L} e_i \frac{G^A_{k,i,t}}{|\langle{r_k}\rangle|}
           + \sum_{i=1}^{L} f_i \frac{G^B_{k,i,t}}{|\langle{r_k}\rangle|}
           + \sum_{i=1}^{23} g_i D_{i,t} + u_t.
    \label{Eq:Pi:IP:Model:LN:PL}
\end{eqnarray}
A comparison of the coefficients of equations (\ref{Eq:Pi:IP:Model:LN:PL}) and (\ref{Eq:Pi:IP:Model:LN}) gives that $a_0^{\rm{ln}}=a_0^{\rm{pl}} + \sum_{i=1}^{L} (c_i^{\rm{pl}}+d_i^{\rm{pl}})$, $a^{\rm{ln}}=a^{\rm{pl}}$, $b^{\rm{ln}}=b^{\rm{pl}}$, $e_i^{\rm{ln}}=e_i^{\rm{pl}}$, $f_i^{\rm{ln}}=f_i^{\rm{pl}}$, and
\begin{equation}
 c_i^{\rm{ln}}=c_i^{\rm{pl}}\beta,  ~~{\mathrm{and}}~~ d_i^{\rm{ln}}=d_i^{\rm{pl}}\beta.
 \label{Eq:ci:di:LN:PL}
\end{equation}
These equations are validated according to table \ref{Tb:Model:LN:L5} and table \ref{Tb:Model:PL:L5}, except for the first one about $a_0$. Particularly, figure \ref{Fig:AllStock:PL:LN} is to verify equation (\ref{Eq:ci:di:LN:PL}). It is found in figure \ref{Fig:AllStock:PL:LN}(a) that the points locate around the dashed diagonal line $y=x$. A linear regression gives that
\begin{equation}
 y = 1.074 x+ 0.003.
 \label{Eq:ci:ln:pl}
\end{equation}
where $y$ contains the data of $c_i^{\rm{ln}}$ and $d_i^{\rm{ln}}$, and $x$ contains the data of $c_i^{\rm{pl}}\beta$ and $d_i^{\rm{pl}}\beta$. To have a better visibility of the points with small values, we plot the absolute values on the log-log scales in figure \ref{Fig:AllStock:PL:LN}(b). We find that several data points for the PS trades deviate the most from $y=x$, where the most deviated point is for $d_3$. The observation of these deviations is consistent with the facts that, for the PS trades, the adjusted R-squares is the smallest and that the associated value of $\beta=0.2$ is not very close to zero.

With the above analysis, we can say that the power-law model and the logarithmic model are in essence equivalent in explaining the immediate price impact.

\section{Conclusion}
\label{S1:Conc}

In this work, we have analyzed the mechanism of immediate price impact by investigating the trade size, the bid-ask spread, the standing volumes of the first few price levels at the bid and ask sides of the limit order book, and the price level gaps. We have proposed two regression models, which have quantitatively similar explanatory powers. We found that large trade sizes, wide bid-ask spreads, high liquidity at the same side and low liquidity at the opposite side will cause large price impacts. For most individual stocks, together with other determinants, the liquidity information at the first two price levels on the limit order book is sufficient to have a high explanatory power. When a cross-sectional regression is performed, it is better to include more price levels. {\color{red}{Asymmetrical impacts have been unveiled for several investigated measures. Especially, the coefficient of spread for FB orders is significant, while that for FS orders is insignificant.}}

More interestingly, we have investigated separately filled orders and partially filled orders. For instance, consider PB orders and FB orders. We found that the coefficient of spread for PB orders is larger than for FB orders ($b_{\rm{PB}}>b_{\rm{FB}}$), the coefficient magnitude of $V_1^A$ for PB orders is smaller than for FB orders ($|c_{1,\rm{PB}}|<|c_{1,\rm{FB}}|$), and the coefficient of $G_1^A$ for PB orders is larger than for FB orders ($e_{1,\rm{PB}}>e_{1,\rm{FB}}$). These observations are consistent with the fact that filled orders are more aggressive than partially filled orders so that liquidity changes have a greater impact on FB orders than PB orders. The results in this work thus extend previous work on the relationship between immediate price impact and order size \cite{Zhou-2012-QF}.

We note that the results in this work are seemingly different from some of previous works \cite{Farmer-Gillemot-Lillo-Mike-Sen-2004-QF,Weber-Rosenow-2006-QF,Joulin-Lefevre-Grunberg-Bouchaud-2008-Wilmott}. This is not surprising since we are analyzing all price impacts, large or small, while those previous works focus on large price changes \cite{Farmer-Gillemot-Lillo-Mike-Sen-2004-QF,Weber-Rosenow-2006-QF,Joulin-Lefevre-Grunberg-Bouchaud-2008-Wilmott}, where the liquidity is found to play a major role and the trade size plays a minor role. In contrast, we showed that both trade size and liquidity (characterized by a measure of market width and two measures of market depth) affect the magnitudes of immediate price impacts. The importance of trade sizes is consistent with earlier works \cite{Lillo-Farmer-Mantegna-2003-Nature,Lim-Coggins-2005-QF,Zhou-2012-QF}. Further more, our analysis here uncovers more determinants of immediate price impacts.

\section*{Acknowledgments:}

This work was partially supported by the National Natural Science Foundation (Grant No. 11075054) and the Fundamental Research Funds for the Central Universities.

\section*{References}

\bibliographystyle{iopart-num} 

\bibliography{E:/papers/Auxiliary/Bibliography}

\end{document}